\documentclass[12pt]{iopart}

\usepackage{iopams}

\usepackage{amsthm,graphicx}

\usepackage{mathrsfs}
\usepackage{amssymb}

\usepackage{lscape,chngpage}

\usepackage{rotating,wasysym}

\def\unwarrow{\nwarrow\joinrel\joinrel\uparrow}

\def\dotsearrow{\hspace{-4mm}\dot{\hspace{4.8mm}\searrow}}
\def\dotswarrow{\hspace{4mm}\dot{\hspace{-4mm}\swarrow}}

\def\be{\begin{equation}}
\def\ee{\end{equation}}
\def\bea{\begin{eqnarray}}
\def\eea{\end{eqnarray}}
\def\bean{\begin{eqnarray*}}
\def\eean{\end{eqnarray*}}

\newtheorem{theorem}{Theorem}

\begin{document}

\article[Singularity theorems]{GR MILESTONE}{The 1965 Penrose singularity theorem}

\author{Jos\'e M M Senovilla}

\address{F\'{\i}sica Te\'orica, Universidad del Pa\'{\i}s Vasco, Apartado 644, 48080 Bilbao, Spain}
\eads{\mailto{josemm.senovilla@ehu.es}}

\author{David Garfinkle}
\address{Dept. of Physics, Oakland University,
Rochester, MI 48309, USA}
\address{and Michigan Center for Theoretical Physics, Randall Laboratory of Physics, University of Michigan, Ann Arbor, MI 48109-1120, USA}
\eads{\mailto{garfinkl@oakland.edu}}

\begin{abstract}
We review the first modern singularity theorem, published by Penrose in 1965. This is the first genuine post-Einstenian result in General Relativity, where the fundamental and fruitful concept of closed trapped surface was introduced. We include historical remarks, an appraisal of the theorem's impact, and relevant current and future work that belongs to its legacy.
\end{abstract}


\section{Introduction}\label{sec:intro}
1965 is an important year in the history of General Relativity (GR), which underwent a very important revival period during the entire 1960 decade. The discovery and interpretation of the Cosmic Background Radiation \cite{PW,DPRW} and the first modern singularity theorem  \cite{P} were published in 1965 ---though both findings happened in 1964--- a year not only right in the middle of that decade, but also exactly 50 years after the birth of Einstein's field equations
\be
R_{\mu\nu} -\frac{1}{2} R g_{\mu\nu} +\Lambda g_{\mu\nu} =\frac{8\pi G}{c^4} T_{\mu\nu} \label{efe}
\ee
hence half way to the present centenary.\footnote{To be historically faithful, version (\ref{efe}) of the field equations only came about one year later when the cosmological constant $\Lambda$ was introduced in \cite{E0}.} In this article we review the history, impact and legacy of the Penrose singularity theorem, one of our selected ``milestones of General Relativity", which was a key contribution to the renaissance of the theory.

The acclaimed singularity theorems are often quoted as one of the greatest theoretical accomplishments in General Relativity, Lorentzian geometry and Mathematical Physics. Even though there were several results proving singularities prior to \cite{P} ---and these will be discussed in sections \ref{sec:prehistory} and \ref{sec:history}--- Penrose's theorem is, without a doubt, the first such theorem in its modern form containing new important ingredients and fruitful ideas that immediately led to many new developments in theoretical relativity, and to devastating physical consequences concerning the origin of the Universe and the collapse of massive stars, see sections \ref{sec:impact} and \ref{sec:numerics}. In particular, Penrose introduced geodesic incompleteness to characterize singularities (see subsection \ref{subsec:singularities}), used the notion of a Cauchy hypersurface (and thereby of global hyperbolicity, see section \ref{sec:theorem}) and, more importantly, he presented the gravitational community with a precious gift in the form of a novel concept: closed trapped surfaces, see subsections \ref{subsec:trapped} and \ref{subsec:trapped2}.

The fundamental, germinal and very fruitful notion of closed trapped surface is a key central idea in the physics of Black Holes, Numerical Relativity, Mathematical Relativity, Cosmology and Gravity Analogues. It has had an enormous influence as explained succinctly in section \ref{sec:impact} and, in its refined contemporary versions --see subsections \ref{subsec:trapped2} and \ref{subsec:ineq}--, keeps generating many more advances (sections \ref{sec:legacy} and \ref{sec:XXI}) of paramount importance and will probably maintain such prolific legacy with some unexpected applications in gravitational physics.

As argued elsewhere \cite{S5}, the singularity theorems constitute the first genuine post-Einstenian content of classical GR, not foreseen in any way by Einstein ---as opposed to many other ``milestones" discussed in this issue.\footnote{As a side historical remark, Einstein himself wrote a paper on ``singularities" \cite{E}, later generalized to higher dimensions with Pauli \cite{EP}. However, these results are concerned with quite another type of 
singularities, namely, with the non-existence of regular point-particle solutions to stationary vacuum gravitational equations ---for a revision see \cite{GG}.} 
The global mathematical developments needed for the singularity theorems, and the ideas on incompleteness or trapping ---and thus also their 
derived inferences --- were not treated nor mentioned, neither directly nor 
indirectly, in any of Einstein's writings. In 1965 GR left adolescence behind, emancipated from its creator, and became a mature physical theory full of vitality and surprises.

\section{Before 1955}\label{sec:prehistory}
Prior to 1965 there were many indications that the appearance of some kind of catastrophic  irregularities, say  ``singularities", was common in GR. We give a succinct summary of some selected and instructive cases, with side historical remarks.
\subsection{Friedman closed models and the de Sitter solution}
As an early example, in 1922 Friedman \cite{Fri} looked for solutions of (\ref{efe}) with the form
\be
ds^2 = -F dt^2 + a^2(t)\left(d\chi^2+\sin^2 \chi d\Omega^2\right) \label{fri}
\ee
where $a(t)$ is a function of time $t$, $F$ is an arbitrary function, and $d\Omega^2$ represents the standard metric for a round sphere of unit radius. This Ansatz followed previous discussions by Einstein and de Sitter \cite{E0,dS} where the space (for each $t=$const.) was taken to be a round 3-sphere (so that $\chi\in (0,\pi)$) and the energy-momentum described pressure-less matter (`dust') with $T_{tt}$ as the unique non-identically vanishing component. Friedman proved that the only possible solutions to (\ref{efe}) with constant $a(t)$ were given by either $F=$const., which gives the Einstein static universe of \cite{E0}, or by $F=c^2 \cos^2\chi$ which leads to the de Sitter universe \cite{dS,dS1} (in this case with $T_{tt}=0$ and $\Lambda >0$). {\it Nevertheless}, he also found that there are dynamical solutions that require $F=c^2$ and a scale factor $a(t)$ satisfying the following set of ODEs
$$
\frac{8\pi G}{c^2}T_{tt} +\Lambda 
=\frac{3}{a^2}(\dot{a}^2+1),\hspace{15mm} 
\Lambda=2\frac{\ddot a}{a}+\frac{1}{a^2}(\dot{a}^2+1)
$$
where dots indicate derivative with respect to $ct$. The second equation has an immediate first integral
$$
a(\dot{a}^2+1)=A+\frac{\Lambda}{3}a^3
$$
for a constant $A$, and then the first equation provides the explicit form of the mass density 
$$
T_{tt} =\frac{3c^2}{8\pi G}\, \frac{A}{a^3}\, .
$$
A simple analysis of the solutions for $a(t)$ leads to the conclusion that, whenever 
$\Lambda <4\pi GT_{tt}/c^2$, $a\rightarrow 0$ inevitably for a finite value of $t$. This is a terrible failure of space-time itself, as the spatial part in (\ref{fri}) truly {\it vanishes} and the mass density diverges. Friedman talked about the ``creation time", and then went on to consider the case with negatively curved space slices $t=$const.\ in \cite{Fri2}.

Incidentally, and for the sake of illustration in later discussions, one may observe that the creation time is absent when $A=0$, in which case the energy density vanishes. This provides another solution of (\ref{efe}) with $a=\lambda \cosh (ct/\lambda)$ and $\Lambda =3\lambda^2$. It is easily checked that this new solution describes a space-time of positive constant curvature, ergo maximally symmetric \cite{Ei,We,MTW}. Surprisingly, the original de Sitter solution with $F=c^2 \cos^2\chi$ has also positive constant curvature, and therefore the two of them  must be isometric. However, there arises a problem in the line-element (\ref{fri}) with $F=c^2 \cos^2\chi$ at $\chi =\pi/2$ ---in this case it is time that seems to vanish--- so that in principle only one of the two intervals $(0,\pi/2)$ or $(\pi/2,\pi)$ should be allowed for $\chi$. This amounts to taking only one hemi-3-sphere, and it becomes apparent that the manifold has been mutilated artificially. 

This problem was addressed in \cite{lanczos,lanczos2} and independently in \cite{weyl}, where one can find pioneer analyses of the observational redshift in cosmological models. With modern notation, the question is resolved by noticing that the first  solution with a constant $a(t)=\lambda$ can be obtained from the second one (written with barred coordinates) by means of the following transformation
$$
t=\frac{\lambda}{c}\ln\left[\frac{\sinh (c\bar t/\lambda)+\cosh (c\bar t/\lambda)\cos\bar\chi}
{\sqrt{1-\cosh^2(c\bar t/\lambda)\sin^2\bar\chi}}\right], \hspace{1cm}
\sin\chi=\cosh (c\bar t/\lambda)\sin\bar\chi \, .
$$
Note that, in this transformation, the barred coordinates are restricted by $\cosh^2(c\bar t/\lambda)\sin^2\bar\chi <1$, a restriction which is not necessary in the solution by itself. Thus, the second solution provides effectively an {\em extension} of the original de Sitter solution. It is important to realize that, while the original de Sitter solution is independent of time $t$ (and therefore it is static), the extended solution depends explicitly on time $\bar t$ and it is not {\em globally} static. One way to see the effect of this property, which is enlightening for later considerations, is to check that the area of the round spheres defined by constant values of $\bar t$ and $\bar \chi$ is given by  $4\pi \lambda^2 \cosh^2(c\bar t/\lambda) \sin^2\bar\chi$. One may be tempted to use this expression as a new coordinate by calling it `$4\pi r^2$' say, but: will $r$ be a space, or rather a time, coordinate? To answer this question one computes the gradient of the function $r$ and checks its causal character. It is easily seen that this gradient is spacelike in the region $\cosh^2(c\bar t/\lambda)\sin^2\bar\chi <1$ (the original de Sitter one), while it is timelike when $\cosh^2(c\bar t/\lambda)\sin^2\bar\chi >1$. Therefore, in this last region, {\em the area of the round spheres is strictly increasing (or decreasing) towards the future, as $r$ is a time coordinate}.

We had to wait until Penrose found a deeper significance and a discerning interpretation of this kind of behavior in his milestone paper \cite{P}.

\subsection{Lema\^{\i}tre: big bang models and an extension of Schwarzschild's solution}
In 1927 Lema\^{\i}tre \cite{Lem} constructed a model that combined Einstein static and de Sitter universes in the sense that it approached the former at large past times and the latter in future distant times. In this way he provided an 
explanation for the observed galactic (or nebulosae) redshift based on GR. Then, in an outstanding and insightful paper \cite{L}, he constructed a general solution of the field equations (\ref{efe}) for dust and spherical symmetry (today known as the Lema\^\i tre-Tolman model \cite{Kra}) and found many interesting results ---including the instability of Einstein static universe---, in particular, an ubiquitous initial singularity of Friedman creation-time type for expanding models capable of explaining the observed cosmological redshifts. Thus, the singularity was there again. Of course, this singular behaviour  could be due to an excess of symmetry (spherical) which, as exact, would not be realistic. Very remarkably he {\em gave up spherical symmetry} and studied the spatially homogeneous but anisotropic models that today we call Bianchi I models \cite{RSh,Exact}. The conclusion was 
unambiguous: the singularity is still there, ``anisotropy can no more prevent the vanishing of space" \cite{L}.

In the very same excellent paper \cite{L}, Lema\^{\i}tre  proved, and unequivocally
understood, the non-singular nature of the Schwarzschild event horizon along the same lines of what had happened with de Sitter space-time and its extensions. He managed to write the general solution of the spherically symmetric vacuum field equations (\ref{efe}) as
$$
ds^2 =-c^2d\tilde{t}^2 +\left(\frac{\alpha}{r}+\frac{\Lambda}{3}r^2\right) c^2 d\chi^2 +r^2 d\Omega^2
$$
where $r^3=\frac{\alpha}{\lambda} \sinh^2[3\lambda c (\tilde{t}-\chi)/2]$, with $\alpha$ a constant and $\lambda$ as above. As one can see, this solution is singular only at $r=0$ (for positive $\alpha$ and non-negative $\Lambda$), thus removing the then called ``Schwarzschild singularity". He then found an explicit change of coordinates bringing the previous line-element into the standard Kottler form \cite{Kot} 
\be
ds^2 = -\left(1-\frac{\alpha}{r}-\frac{\Lambda}{3} r^2\right)c^2 d\hat t^2
+\left(1-\frac{\alpha}{r}-\frac{\Lambda}{3} r^2\right)^{-1}dr^2 +r^2 d\Omega^2 . \label{kot}
\ee
Recall that Schwarzschild had found the general solution of Einstein vacuum equations without $\Lambda$ in \cite{Sch}, which can be written in standard coordinates as (\ref{kot}) with $\Lambda =0$ \cite{Sch,Droste}. There appeared a distinguished worrying hypersurface $r=\alpha$, which is actually similar to the region $\chi =\pi/2$ in de Sitter space-time analyzed above. Therefore, the metric found by Lema\^{\i}tre is an explicit regular extension of the Schwarzschild solution that includes regions beyond $r=\alpha$. In \cite{L} he
distinctly remarked that the ``problem'' was due to the assumption
of the entire spherically symmetric world being static: 
\begin{quotation}
{\em We show that the [$r = \alpha$] singularity of the Schwarzschild exterior is an
apparent singularity due to the fact that one has imposed a static solution and
that it can be eliminated by a change of coordinates.}
\end{quotation}
And later
\begin{quotation}
{\em The [$r = \alpha$] singularity of the Schwarzschild field is thus a fictitious
singularity, analogous to that which appears at the horizon of the centre in the
original form of the de Sitter universe.}
\end{quotation}
The question of why, after the resolution of the de Sitter and Schwarzschild horizons, the confusion about the
latter went on for almost another 40 years ---see e.g. the
excellent reviews in \cite{TCE,I}--- is a mysterious story. Actually,  Eddington had found another extension (what we would call today its Kerr-Schild form) in 1924 \cite{Edd}. This is the basis of what was later called the Eddington-Finkelstein extension \cite{Fin} ---explicitly used for the illustrated discussion in our milestone paper \cite{P}--- and then the maximal Kruskal extension \cite{Kru}. As clearly discussed in \cite{Nov} where the so-called $R$-and $T$-regions were introduced, all extensions require giving up staticity if regions with $r<\alpha$ are to be included, where $4\pi r^2$ is the area of the preferred round spheres. These extensions show that there is something unusual going on with the round spheres in the $r<\alpha$-regions because their area function $4\pi r^2$ can be seen as a {\it timelike} coordinate there, as mentioned above in the de Sitter case.

It is interesting to notice that Lema\^{\i}tre wanted to solve an apparent
contradiction between the spherically symmetric Friedman's solutions, where the area of the round spheres 
can become as small as desired, and the existence of a minimum value for such an
area if one is to accept the $r>\alpha$ restriction of the static Schwarzschild solution because then ``a mass such as that of the Universe [could] not have a radius less than one billion light years'' (\cite{L}, p.643 of the english translation).
There is a discussion in \cite{E93} putting forward the view that this line
of reasoning was seminal to the later Oppenheimer-Snyder models, next subsection \ref{subsec:OS}.

\subsection{The Oppenheimer-Snyder model}\label{subsec:OS}
The previous discussion indicates that singularities in the past of our world appeared in the simplest models of the classical GR theory if the Universe is expanding everywhere. One could consider a sort of time-reversal of this situation: what happens in contracting worlds? This turned out to be of enormous relevance for the study of compact stars, since in 1931 Chandrasekhar unexpectedly found an upper mass limit for white dwarf stars in 
equilibrium, even when taking into account the quantum effects \cite{Chandra}. This implied that
stars with a larger mass will inevitably collapse. Then, the question of massive neutron cores (or stars) was addressed in \cite{OV} by using a cold Fermi gas equation of state and GR. They found another mass limit for equilibrium and concluded that, even allowing for deviations from the Fermi equation of state, a massive enough neutron star will contract indefinitely never reaching equilibrium again. 

This prompted Oppenheimer and Snyder to consider the solutions of the field equations (\ref{efe}) that described such physical processes \cite{OS}. They proved using general arguments that, in spherical symmetry,  values of $r=\alpha$ would eventually be reached, that light emitted from the star would be more and more redshifted for external observers ---who would only see the star approach $r\rightarrow \alpha$ asymptotically---, and that the entire process will last a finite amount of time for observers comoving with the stellar matter. They then constructed an explicit analytical model which, in modern language, consists of a portion of the Friedman closed model (\ref{fri}) for dust (with $\Lambda =0$ and $\dot a <0$) matched with the Schwarzschild solution at the timelike hypersurface defined by $\chi =\chi_0 <\pi/2$ on the interior side ---and correspondingly by a hypersurface ruled by timelike geodesics and $r=a(t) \sin\chi_0$ on the vacuum side--- proving that the junction requires 
$$
\alpha = A \sin^3\chi_0.
$$ 
Hence, (i) the ``Schwarzschild surface" $r=\alpha$ was indeed crossable by innocuous models 
containing realistic matter such as dust; and (ii) a careful analysis of the  model shows that the star will end up in a catastrophic singularity where $a(t)\rightarrow 0$ and therefore space ``vanishes" again.

What did Einstein think about the singularities and all the previous results? Well, it is hard to tell, obviously, but it seems that he ---and the orthodoxy--- simply dismissed the known singularities as either a mathematical artifact due to the spherical symmetry, or as unattainable effects beyond the feasibility of the physical world. In the same 1939 Einstein, seemingly unaware of \cite{OS}\footnote{\cite{OS} was published in september, before \cite{E2} (october). However, Einstein's paper was submitted before (may vs. july).}, ``proved" that the Schwarzschild singularity is not physically realizable  \cite{E2} by considering a statistical distribution of particles moving in circular orbits due to its own (spherically symmetric) gravitational field. He concluded that $r\rightarrow \alpha$ cannot be reached for the reason that matter cannot be concentrated arbitrarily as otherwise the constituting particles would reach the speed of light ---and he conjectured that this behaviour will be exhibited in all cases. Then, in \cite{ESt} he treated the problem of the influence of the universal expansion in local gravity fields and came up with a solution widely known as the Einstein-Straus vacuole, which is the basis of the Swiss-cheese models, by matching an interior Schwarzschild static cavity to an external Friedman-Lema\^\i tre expanding universe ---for a recent review, see \cite{MMV}. Ironically, this is {\em precisely} the same matching as that in \cite{OS} ---for closed universes and in time reversal. Whenever two portions of given spacetimes are matched across corresponding proper hypersurfaces, the two discarded pieces in the given spacetimes also automatically match with exactly the same conditions: these are called ``complementary matchings", see e.g. \cite{FST}. Accordingly, the Einstein-Straus model, when contemplated in its full {\em past} evolution, leads to a singularity and to regions with $r<\alpha$ in the vacuum part of the space-time.

In summary, even though infinite values of 
physical observables must not be accepted in physical reality, one must be 
prepared to probe the limits of any particular theory, but this was reluctantly done in GR before 1955 despite many important indications that this was needed. A new generation of less prejudiced physicists and mathematicians 
was about to enter into play to take the question 
of the singularities seriously within the GR theory. As a final remark of historical importance and scientific relevance, let us mention that  G\"odel wrote his famous paper \cite{Go} with the solution named after him in a volume dedicated to 
Einstein's 70th birthday. It is a ``totally vicious" solution of the field equations (\ref{efe}), meaning that there are closed timelike curves passing through every point of the manifold \cite{Ca,S1}. Nevertheless, G\"odel's space-time is geodesically complete, free of singularities, and rotating. He studied in further depth the case of rotating universes in a second paper \cite{Go1}. These papers are considered 
\cite{TCE} the genesis of many of the necessary 
techniques used in the path to the singularity theorems, specially concerning causality theory and actions of Lie groups on the space-time, see \cite{Ell}. It may have also influenced Raychaudhuri in the formulation of his fundamentally important equation, see next section.

\section{From 1955 to 1965}\label{sec:history}
Einstein died on 18th April 1955 and, coincidentally, less than one month later a new era started for GR with the publication of Raychadhuri's paper containing the first ever singularity theorem \cite{Ray}. In this remarkable paper, he 
included a primitive form of the equation named after him,  which is the basis of later developments and of
{\em all} the singularity theorems, see subsection \ref{subsec:thms}. 

The Raychaudhuri equation can be readily derived from the Ricci identity for a vector field $u^\mu$
$$
(\nabla_{\mu}\nabla_{\nu}-\nabla_{\nu}\nabla_{\mu})u^{\alpha}=
R^{\alpha}{}_{\rho\mu\nu}u^{\rho}
$$
by contracting $\alpha$ with $\mu$ and then with $u^{\nu}$, leading to
$$
u^{\nu}\nabla_{\mu}\nabla_{\nu}u^{\mu}-u^{\nu}\nabla_{\nu}\nabla_{\mu}u^{\mu}=
R_{\rho\nu}u^{\rho}u^{\nu}
$$
and then reorganizing by parts the 
first summand on the left-hand side:
\begin{equation}
u^{\nu}\nabla_{\nu}\nabla_{\mu}u^{\mu}+\nabla_{\mu}u_{\nu}\nabla^{\nu}u^{\mu}-
\nabla_{\mu}(u^{\nu}\nabla_{\nu}u^{\mu})+R_{\rho\nu}u^{\rho}u^{\nu}=0.
\label{rayeq}
\end{equation}
Nothing deep or mysterious here: Raychaudhuri's
contribution was to understand the deep
physical implications of this simple geometrical relation. 
To start with, note that if $u^\mu$ defines a 
(affinely parametrized) {\em geodesic} vector field, then 
$u^{\nu}\nabla_{\nu}u^{\mu}=0$ and the third term in (\ref{rayeq}) vanishes. The 
second term can be rewritten by splitting 
$$
\nabla_{\mu}u_{\nu}=\nabla_{(\mu}u_{\nu)}+\nabla_{[\mu}u_{\nu]}\equiv S_{\mu\nu}+A_{\mu\nu}
$$
into its symmetric $S_{\mu\nu}$ and antisymmetric $A_{\mu\nu}$ parts, 
so that
$$
\nabla_{\mu}u_{\nu}\nabla^{\nu}u^{\mu}=S_{\mu\nu}S^{\mu\nu}-A_{\mu\nu}A^{\mu\nu}\, .
$$
A one-form $u_{\mu}$ is proportional to an exact differential $u_\mu \propto \partial_\mu f$ ---defining thereby  orthogonal 
hypersurfaces $f=$const.--- if and only if $A_{\mu\nu}=0$. Moreover,
assuming that $u^{\mu}$ is either null or timelike normalized, so that $u^\mu u_\mu =0$ or $-1$, both $S_{\mu\nu}S^{\mu\nu}$ and $A_{\mu\nu}A^{\mu\nu}$ are non-negative.  Consequently, for 
hypersurface-orthogonal geodesic time-like or null vector fields 
$u^{\mu}$, one has
$$
u^{\nu}\nabla_{\nu}\nabla_{\mu}u^{\mu}=
-S_{\mu\nu}S^{\mu\nu}-R_{\rho\nu}u^{\rho}u^{\nu},
$$
ergo the sign of the derivative of the divergence or {\em expansion}
\be
\theta\equiv \nabla_{\mu}u^{\mu}=S^{\mu}{}_{\mu} \label{theta}
\ee
along the geodesic congruence
is determined by that 
of $R_{\rho\nu}u^{\rho}u^{\nu}$. If the latter is non-negative, 
the former is non-positive. In particular, if $\theta$ is 
negative (positive) at some point and $R_{\rho\nu}u^{\rho}u^{\nu}\geq 0$ it follows that $\theta$ will reach an infinite negative value in finite affine parameter to the future (past) ---unless all the quantities are zero 
everywhere. If (a timelike)  $u^\mu$ describes the motion of a fluid moving along these geodesics, then 
a {\em physical} singularity develops, since the mean volume 
decreases and the density of the fluid will become unbounded, see subsection \ref{subsec:Ray-Kom}. This was  
the situation treated in \cite{Ray} for the case of irrotational 
dust. 

In general, though, no singularity but rather a {\em caustic} along the flow lines of the congruence 
defined by $u^{\mu}$ is formed. This property is usually called the 
{\em focusing effect} on causal geodesics. For this to take place one needs the condition 
\begin{equation}
R_{\rho\nu}u^{\rho}u^{\nu}\geq 0 \label{sec}
\end{equation}
which is a {\em geometric} condition independent of the particular theory. 
(\ref{sec}) is called the {\em timelike (respectively null) convergence condition} when valid for timelike (resp. null) vector fields. In GR, of course, one can relate the 
Ricci tensor to $T_{\mu\nu}$ via (\ref{efe}) and thereby condition (\ref{sec}) can be 
rewritten in terms of physical quantities. This is why in the standard literature (\ref{sec})  is many times called the {\em strong energy condition} in the case with $\Lambda =0$ \cite{HE}, see subsection \ref{subsec:sec} for a discussion.

\subsection{The Raychaudhuri and Komar singularity theorems}\label{subsec:Ray-Kom}
The main result found in \cite{Ray} can be stated as a singularity theorem for irrotational dust. This was almost immediately generalized, independently, by Komar to the general situation of fluids \cite{K,Ray2}. Komar's paper contained basically
the same ideas but the concept of an energy
condition on $T_{\mu\nu}$ ---such as (\ref{sec}) when (\ref{efe}) is taken into account--- was introduced. {\it Assuming} that matter moves along a geodesic and hypersurface orthogonal timelike vector field $u^\mu$ a matter singularity can be obtained under some physically interesting circumstances. In modern terms the theorem can be stated as (Theorem 5.1 in \cite{S1}):
\begin{theorem}[Raychaudhuri and Komar]
Assume $\Lambda =0$ and a perfect-fluid energy-momentum tensor
\be
T_{\mu\nu}=\varrho u_{\mu}u_{\nu}+p(g_{\mu\nu}+u_\mu u_\nu), \hspace{1cm} u^\mu u_\mu =-1
\ee
whose velocity vector field 
$u^\mu$ is geodesic and irrotational. If the expansion (\ref{theta}) is
positive (resp. negative) at an instant of time and (\ref{sec})  holds, then the energy density $\varrho$ of the fluid diverges in
the finite past (future) along every integral curve of $u^\mu$.\label{th1}
\end{theorem}
The notion of `an instant of time' is meaningful here:  as $u^\mu$ is geodesic and hypersurface-orthogonal we have
$u_\mu =-\partial_\mu \tau$, where $\tau$ is a natural time coordinate for the fluid, and the assumption reads simply $\theta |_{\tau_0}>0$ (or $<0$). The key assumption in Theorem \ref{th1} is the absence of
acceleration and rotation, from where we learn that acceleration (or
rotation) of matter becomes necessary to avoid singularities. This is 
physically reasonable for (i) acceleration is directly related to the existence of gradients of pressure, which are forces
opposing gravitational attraction in general fluids; and (ii) G\" odel's revolutionary papers \cite{Go,Go1} proved that rotation could prevent the formation of singularities.  This is also
supported by Newtonian cosmologies, in which rotation prevents the
appearance of matter singularities, see, e.g., \cite{RSh} and references
therein. 

Theorem \ref{th1} has one virtue that, as we will see in subsection \ref{subsec:thms}, will be lost in the more advanced singularity theorems: it
predicts that the singularity means a divergence of the energy density, and it says where to locate it.

Approaching 1965, there were more singularity theorems based on equation (\ref{rayeq}), mainly dealing with spatially homogeneous (Bianchi) cosmologies \cite{Sh0,HE2,LK,LK1}. In particular, \cite{LK,LK1} were the basis for the later much studied BKL conjecture \cite{BKL,BKL1}, see subsection \ref{subsec:BKL}.

To finish this section, we would like to remark that 
in 1963 Kerr also discovered the first solution for a spinning mass \cite{Kerr} later to become the unique metric of uncharged black holes \cite{Rob,Heus}.

\section{The 1965 theorem, its implications and relevance}\label{sec:theorem}
The focusing effect predicted by the Raychaudhuri equation (\ref{rayeq}) on causal geodesics emitted orthogonally from co-dimension 1 (in the case of hypersurface-orthogonal timelike unit vector fields $u^\mu$) or from co-dimension 2 (null vector field case) submanifolds became a fundamental ingredient to derive the 
powerful singularity theorems. Nevertheless, as remarked above, this 
focusing by itself does not lead to singularities in general. As a 
trivial example, observe that flat spacetime satisfies condition 
(\ref{sec}) trivially, but it has no singularities: the focusing 
effect simply leads to focal points or caustics of the geodesic 
congruences. 

One therefore needs to combine the focusing of geodesics with other reasonable ingredients. In this respect, the key turning point was the 1965 singularity theorem \cite{P}. As stated in the introduction of \cite{P}, Penrose wanted to prove that deviations from spherical symmetry were not able to prevent the formation of singularities, such as those described in the Oppenheimer-Snyder collapse, within the GR theory. Recall that there was a previous result by Lema\^\i tre  in this direction, but it still needed a lot of symmetry (spatial homogeneity); work in \cite{DZN} also led to the conclusion that irregularities in non-spherical collapse were somehow suppressed by physical processes leading to a picture very similar to that of \cite{OS}. To achieve his goal, Penrose brought in the idea of {\em incompleteness} to describe singular spacetimes (subsection \ref{subsec:singularities}), and he introduced the notion of {\em closed trapped surface} for the first time, a major conceptual contribution to the physics of the gravitational field (subsection \ref{subsec:trapped}). Before discussing  these two fundamental ideas  let us present a modern version of Penrose's theorem.

\begin{theorem}[Penrose singularity theorem]\label{th:P}
If the space-time contains a non-compact Cauchy hypersurface $\Sigma$ and a closed future-trapped surface, and if the convergence condition (\ref{sec}) holds for null $u^\mu$, then there are future incomplete null geodesics.
\end{theorem}

For a proof of the theorem one can consult many references apart from \cite{P}, e.g. \cite{HE,P5,S1,Kri,BE,O,Wald}. 

A Cauchy hypersurface \cite{HE,Ge3,P5,Wald} is an ``instant of time'' that provides good initial value conditions for the {\em entire} space-time. More precisely, it is an achronal hypersurface which is met once and only once by all causal endless curves (actually this follows if it is traversed by all endless null geodesics, \cite{Ge3}). This is the pertinent property for the proof of theorem \ref{th:P}, but there are other implications of the existence of a Cauchy hypersurface which will be essential for later singularity theorems, namely, that this is equivalent to the space-time being globally hyperbolic \cite{Le}, and the existence of maximal geodesics between any two causally related points, see subsection \ref{subsec:causalcond} for details. If there is a Cauchy hypersurface $\Sigma$, then the space-time as a manifold is a product $\mathbb{R}\times \Sigma$, all slices $\{t\}\times \Sigma$ being diffeomorphic Cauchy hypersurfaces \cite{Ge3,Se,BS}. The presence of a Cauchy hypersurface is not guaranteed as was shown by Penrose in another influential paper \cite{P2} ---published almost simultaneously with \cite{P}--- where he discovered that plane waves in GR do not admit Cauchy hypersurfaces. Whether or not physically realistic spacetimes are globally hyperbolic is related to the so-called {\em strong cosmic censorship} conjecture, see subsection \ref{subsec:CC}.

The strategy of the sketched proof presented in \cite{P} was to assume that null geodesics were complete, proving that then the boundary of the future of the closed trapped surface is compact. To do that one uses the focusing of null geodesics emanating orthogonally from the surface as follows from (\ref{rayeq})\footnote{Penrose used an equation from \cite{NP} and did not refer to Raychaudhuri's equation. However, in his previous paper \cite{P2} he acknowledges that this effect is ``essentially the same phenomenon as that discovered" in \cite{Ray,K}, applied to null rather than timelike geodesics.}. But this boundary is actually an imbedded submanifold (without boundary!) \cite{HE,P5}, while its canonical projection to $\Sigma$ would have to have a boundary as $\Sigma$ is non-compact. Therefore, as argued in \cite{P}, one must choose among one of the following possibilities: (\ref{sec}) is violated for null vectors, or the concept of space-time loses its meaning at extreme conditions (maybe quantum effects), or some null geodesics are not complete. Actually, these may all be inter-related.

Next, we discuss the idea of incompleteness, which under the influence of \cite{P} eventually became the standard definition of a singularity in GR.

\subsection{What is a singularity?}\label{subsec:singularities}
The problem of how to define a singularity in 
General Relativity was very difficult indeed, see \cite{Ge2,RSh}.
Intuitively, one expects that divergences of any physical or 
geometrical quantity would be a characteristic feature of singularities. 
However, there are problems of several kinds: 
\begin{enumerate}
\item singularities do not belong to the space-time which is by definition constituted of regular points. 
\item characterizing the singularities with misbehaviors of the curvature tensors may run into problems,
as they may depend on a bad choice of basis  \cite{ES}, 
\item even if one uses only curvature invariants ---independent 
of the bases--- it can happen that all of them vanish and still there 
are singularities.
\item sometimes, singularities arise due to bad properties of the tangent bundle, for instance in conical singularities the main problem is a lack of independent tangent directions \cite{ES,VS}
\item there are other known, more pathological, examples of spacetimes with vanishing curvature and incomplete geodesics \cite{Mis2,Mis}
\item sometimes there appear {\em directional} singularities, defined as limit points towards which the curvature tensor blows up along some, but not along other, directions \cite{St,GA,Ho,SS}
\end{enumerate}
These complications led to an elaborated classification of possible singularities arising from the curvature tensors \cite{ES}.

On the other hand, incomplete physical curves may help here. Curves are very good pointers, and using curves one  employes objects that belong to the space-time exclusively. One can imagine the fate of a courageous traveler (his/her worldline is a casual curve in the space-time) 
approaching a singularity: he/she will disappear from our world in a {\em finite} time. 
 The time-reversal situation will describe 
the ``creation'' of the Universe: things suddenly appeared from 
nowhere a {\em finite} time ago. 
It seems thus sensible  to
diagnose the existence of singularities whenever there are (hypothetical) travelers 
which disappear, or materialize, abruptly.

\begin{quotation}
{\it A singularity in a Lorentzian manifold  is an incomplete endless curve.}
\end{quotation}

These curves cannot be continued regularly within the 
space-time despite they reach only finite values of their canonical 
parameter. All singularity theorems after \cite{P} prove merely the existence of {\em geodesic} incompleteness, 
which of course is a {\em sufficient} condition for singular spacetimes according to the definition. Nevertheless, there are known examples \cite{Ge2,Beem} of geodesically complete space-times 
with incomplete time-like curves of everywhere bounded acceleration, ergo singular too. As far as we know, there is no known singularity theorem proving the existence of a geodesically complete singular space-time of this type.

What is the relation between geodesic incompleteness 
and curvature misbehavior, if any? 
On can actually prove limits on the curvature growth for maximal geodesics
\cite{T2,KR,Ne,Szab}. The main result is that, along a causal incomplete geodesic with tangent vector $v^{\mu}$, $R_{\alpha\beta\mu\nu}v^{\beta}v^{\mu}$ computed in a parallel propagated basis cannot grow faster, in modulus, than
$(\tau -\hat{\tau})^{-2}$ when approaching the singularity at $\tau = \hat{\tau}$, where $\tau$ is an affine parameter (proper time for timelike geodesics) along the geodesic.
It is certainly curious that one can put a limit on the curvature growth when
approaching the end of an incomplete geodesic predicted by the theorems, but one
does not know whether the curvature will diverge at all! However, some results pointing towards a curvature divergence can be found in \cite{Cla1,Cla2,Cla3,Cla4,CS,Tho,TCE}.

\subsection{The key concept of closed trapped surface}\label{subsec:trapped}
In Newtonian gravitation there is the important concept of escape velocity, providing the necessary initial condition for a small object to abandon a gravitational field eventually reaching infinity. Thus, a good indication of a very strong gravitational field would be a very large escape velocity ---say close to that of light. In GR things are much more complicated, as one would expect. Strong gravitational fields can distort the paths of light immensely. Penrose cleverly invented an appropriate notion, {\em closed trapped surfaces}, capturing this intuitive but maybe vague idea of ``possibility of getting away from a gravity field".

\begin{figure}
  \includegraphics[height=.35\textheight]{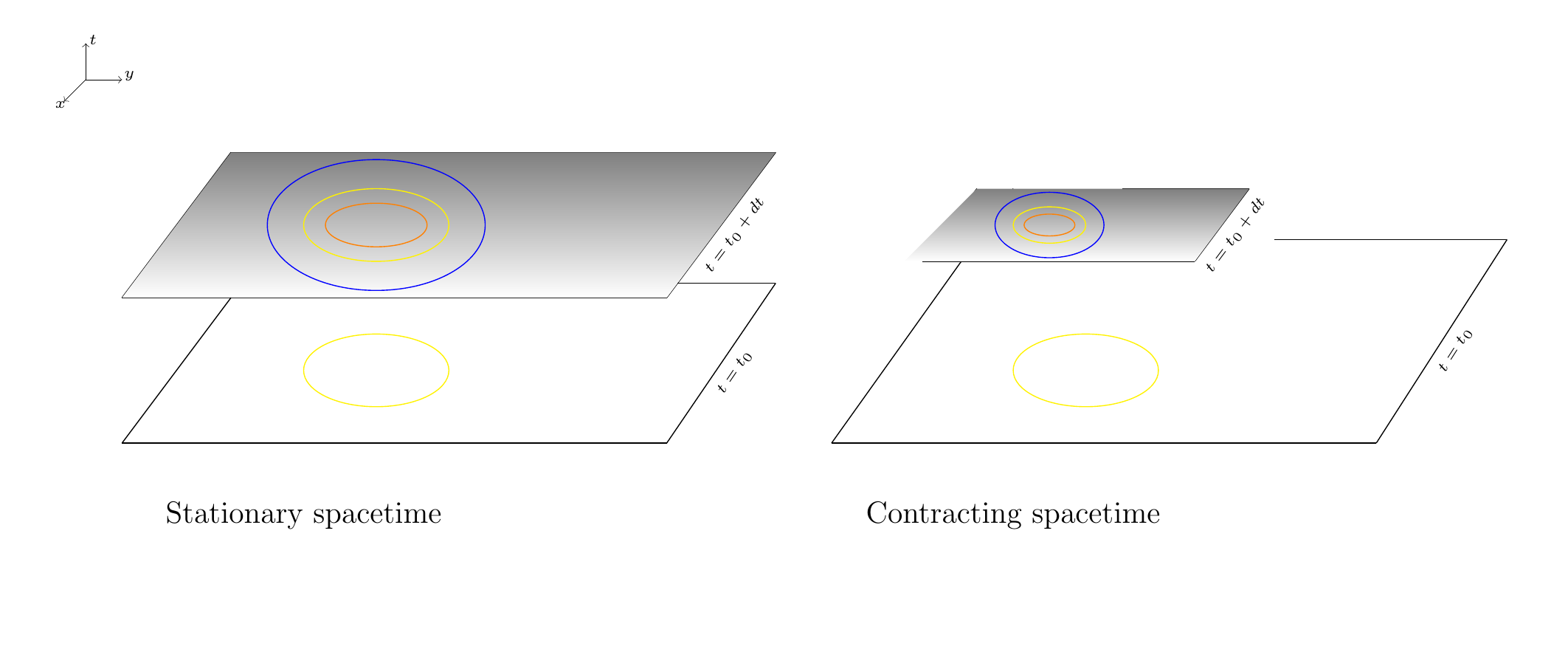}
  \caption{Intuitive explanation of the concept of a closed future trapped surface, one dimension suppresed. Time is running upwards as shown. In the stationary situation (left), the original sphere represented in the $t=t_0$ slice as a yellow circle sends light signals in the radial out- and in-going directions. These flashes form two new spheres after an infinitesimal time $dt$, as represented by the red (ingoing) and blue (outgoing) circles. The area of the former (latter) is smaller (larger) than the area of the orginal surface, which remains constant with time $t$. However, if the universe is contracting, as shown on the right, both the blue and red surfaces can have areas smaller than that of the original surface {\em at the initial time}. In this case, the surface is said to be future trapped. The limit case where the blue surface has precisely the area of the original surface at the initial time is termed {\em marginally} future trapped. (Reproduced from \cite{SMex})}
  \label{fig:Senovilla1}
\end{figure}

As introduced in \cite{P}, a closed trapped surface is a two-dimensional imbedded submanifold $S$ (surface), compact without boundary (closed), such that the two families of light rays emerging orthogonally from $S$ towards the future converge initially (trapped). An intuitive picture of this idea is presented in figure \ref{fig:Senovilla1}.
Locally, the two family of light rays can be mathematically represented by a pair of future-directed affine-parametrized geodesic null vector fields $k_\pm^\mu$, and the infinitesimal variation of the area of $S$ is measured by the respective expansions (\ref{theta}): $\theta_\pm =\nabla_\mu k^\mu_\pm$. These are defined up to multiplicative factors, but only their signs are relevant here. Thus, a closed future-trapped surface has
\be
\theta_+ <0, \hspace{1cm} \theta_- <0 . \label{trapped}
\ee
There is a dual definition of {\em past}-trapped surface replacing future by past or, equivalently, keeping the future-pointing $k^\mu_\pm$ but reversing the signs in (\ref{trapped}).

The notion of trapped surface is independent of the coordinates and of the existence of symmetries such as spherical or the axial symmetry present in Kerr's solution \cite{Kerr,Exact,HE}. Importantly, it is defined by using {\em inequalities} (\ref{trapped}), so that trapped surfaces remain stable under small perturbations; this can be made mathematically precise within the space of Lorentzian metrics \cite{Ler}. It is therefore a general concept devised to represent finite regions (the interior of the surface) which are somehow instantaneously fully confined by a gravitational field.

It is easy to check that (\ref{trapped}) is precisely the mathematical condition which generalizes the timelike character of the gradient of $4\pi r^2$ as it appeared in the earlier extensions of de Sitter and of Schwarzschild spacetimes (for $r<\alpha$), as discussed above. For, if the gradient of $r$ (which measures the instantaneous variation of area) is timelike, its scalar product with a future-pointing null vector is negative (if $dr$ is future), or positive (if $dr$ is past). But these scalar products are proportional to the expansions $\theta_\pm$. In other words, condition (\ref{trapped}) is equivalent to $dr$ being timelike future (and past for the reversed signs). Observe that actually this means that that the scalar product of $dr$ with {\em any} future-pointing vector has the corresponding sign. Hence, a future (past) trapped surface has an area that is initially decreasing along {\em any} possible future (past) direction. Two important remarks:
\begin{itemize}
\item the one-form $dr$ (``gradient of area"), and its causal character, emerges as the key object to define the trapping of surfaces in known situations with spherical symmetry. We will see that this can be readily generalized, and the correct mathematical vector to be used is simply the {\em mean curvature vector} of $S$ \cite{Kri,MS,O}.
\item From the discussion on de Sitter and Schwarzschild extensions in section \ref{sec:prehistory}, and from the above definition and intuitive picture of closed trapped surfaces, one realizes that {\em stationary} regions cannot accommodate them. This is actually a general result, see \cite{MS}.
\end{itemize}
The idea of closed trapped surface and its applications will be analyzed in deeper detail, with a more up-to-date perspective, in subsection \ref{subsec:trapped2}.

\section{After 1965: immediate impact of the theorem}\label{sec:impact}
Penrose's  singularity theorem shook the GR community. Its impact was ample, profound and straightaway, see \cite{Ell1,TCE}.

A few months after \cite{P} was published, Hawking \cite{H0} realized that closed trapped surfaces, in its {\em past} version, will be present in any expanding Universe close to be spatially homogeneous and isotropic. This started a series of papers by him, Ellis, Geroch and others on the question of the inevitability of an initial singularity in our past if GR is assumed to hold and some reasonable conditions are met. Papers were even published in the same issue and consecutively, such as \cite{H00,Ge-}. 

In \cite{H1,H2,H3} Hawking developed new ideas which ---despite some initial inaccuracies \cite{TCE}---, when combined with those put forward in \cite{P}, formed the nucleus of the modern singularity theorems. In particular he borrowed from Riemannian geometry the concept of focal and conjugate points (essentially, these are the caustics predicted by (\ref{rayeq})) and their significance explained by means of the first and second variation of the length integral \cite{H1}: the outcome is that a timelike curve from a point (respectively from a spacelike hypersurface $\Sigma$) maximizes the arc-length if and only if it is a geodesic (orthogonal to $\Sigma$), without corners and without conjugate (resp. focal) points, see \cite{BE,O,HE,Kri,S1,Wald}. Similar results, adequately adapted, hold for null geodesics from a point, or orthogonal to a spacelike surface. A theorem was proven by assuming a Cauchy hypersurface $\Sigma$, letting $u^\mu$ be the unit geodesic vector field orthogonal to $\Sigma$ on $\Sigma$, and assuming that its expansion (\ref{theta}) is bounded from below by a positive constant. The Raychaudhuri equation leads then to all the geodesics tangent to $u^\mu$ having a finite length to the past, ergo the singularity \cite{H1,HE,S1}. This mainly applied for non-compact $\Sigma$, the compact case was treated in \cite{H2}. In \cite{H3} the previous theorem was improved and another theorem was found by assuming the existence of a point with re-converging light cones towards the past \cite{H3}. This is much better adapted to what we know about the Universe (no need to assume an everywhere expanding slice), and the theorem was used in an important article \cite{HE3} to show that a combination of the two 1965 breakthroughs ---cosmic background radiation and singularity theorems--- gave strong indication that a singularity in our past was inevitable if one is to assume GR and condition (\ref{sec}) holds everywhere.

Many other physical and mathematical developments were catalyzed by \cite{P} and the papers cited in the previous paragraph. For instance, in \cite{dW} the recurrent idea of whether or not a quantum theory of gravity can extend solutions of classical GR beyond the singularities was analyzed for the first time. Furthermore, the theory of causality, and of causality conditions, was developed and has become an integral part  of GR. The foundations were established in \cite{KrP} for an abstract analysis of causal spaces, then the stable causality condition and time functions were introduced in \cite{H}, the relation between the topology of spatial sections and causality studied in \cite{Ge}, the basic properties of domain of dependence and global hyperbolicity given in \cite{Ge3}, and a comprehensive summary provided in \cite{Ca}, see also \cite{Ge4}. Updated  reviews are \cite{GS,MSa}.

Another important ramification from the ideas behind the singularity theorems was the standard theory of black holes \cite{H4}. It all started with another forceful paper by Penrose \cite{P0} where the fecund idea of conformal infinity was put forward, leading to a definition of (weak) asymptotic flatness \cite{P3} and thereby to the notion of event horizon: the boundary of the past of future null infinity, see \cite{Ge4,HE,P5,PR1,PR2} and for a recent review of conformal infinity \cite{Fra}.

Eventually, a considerable improvement in the understanding of the conditions under which a space-time must be singular was accomplished by Hawking and Penrose \cite{HP} in what is still considered the preeminent singularity theorem. Later, excellent condensations of the techniques and results behind the theorems, with lengthy enlightening discussions, were given in \cite{HE,P5}. The Hawking-Penrose result is based on the fundamental observation that the following three things are incompatible: 
\begin{itemize}
\item every endless causal geodesic has conjugate points 
\item there are no closed timelike curves
\item there is a set whose future (or past) has a compact boundary
\end{itemize}
These last sets are called ``trapped sets'' in the literature \cite{HE,HP,S1}, but they are not to be confused with trapped surfaces. The concept of trapped surface is local, there is no need to know the structure of its future or past. However, one can prove, following the ideas in \cite{P}, that a closed trapped surface will become a trapped set whenever the space-time possesses complete null geodesics \cite{BE,S1}. The same can be said of points with reconverging light cones, and of compact imbedded hypersurfaces. To enforce the first of the three conditions usually (\ref{sec}) is enough, but to avoid exceptional situations one also needs to add the so-called {\em generic}  condition: 
\be
u_{[\rho}R_{\alpha]\beta\lambda[\mu}u_{\sigma]}u^\beta u^\lambda \neq 0.\label{gen}
\ee
For timelike $u^\mu$ this can be written in the simpler form $R_{\alpha\beta\lambda\mu}u^\beta u^\lambda \neq 0$.
Physically this means that the tidal force felt by causal curves will not always and everywhere be aligned with their tangent vectors. Of course, there are spacetimes where this condition will not hold, but they are very special indeed, see subsection \ref{subsec:sec}.

As a corollary one has the theorem
\begin{theorem}[Hawking and Penrose]\label{th:HP}
If the convergence (\ref{sec}) and generic (\ref{gen}) conditions hold for causal vectors, there are no closed timelike curves and there exists at least one of the following:
\begin{itemize}
\item a closed achronal imbedded hypersurface
\item a closed trapped surface,
\item a point with re-converging light cone
\end{itemize}
then the space-time has incomplete causal geodesics.
\end{theorem}
Even though originally proven in 4-dimensional spacetimes, Theorem \ref{th:HP} actually holds in arbitrary dimension ---in which case the closed trapped surface is a co-dimension two trapped submaniolfd.

In the next subsection we present an assessment of this and other classical singularity theorems, the ideas behind their proofs, and a discussion of their assumptions.

\subsection{Classical singularity theorems}\label{subsec:thms}
Since the publication of \cite{P,HP} there have been many papers proving singularity theorems, mainly trying to relax the assumptions, to refine their conclusions and/or to enhance their possible physical implications. Letting aside some 
subtleties, they are usually interpreted as providing evidence of the (classical)
singular beginning of the Universe and of the singular final fate of massive compact
stars. A more up-to-date perspective would rather state that they provide very solid evidence of the need of (possibly quantum) corrections to GR when extreme gravity effects occur.

Some selected improved theorems are: 
\begin{itemize}
\item adding alternatives to the trapped sets in theorem \ref{th:HP} and some topological considerations for the case of non-simply connected space-times \cite{Gan,Gan2,Lee}
\item relaxing (\ref{sec}) to some milder averaged condition along geodesics, e.g.  \cite{T6,ChE,Bor,KR}
\item related to the previous, theorems which allow for the violation of (\ref{sec}) to accommodate inflationary models \cite{Bor2,BV,BV2,BV3} (more on this in section \ref{sec:XXI})
\item quantifying somehow the curvature growth along incomplete geodesics and proving the existence of a {\it maximal} incomplete geodesic, that is, one maximizing proper time between any two of its points \cite{Szab,KR}
\item weakening the causality assumption that forbids closed timelike curves  \cite{T,T4}. This
work was later improved in \cite{Kr} assuming that the boundary of the set of
points where there are closed causal curves is compact, see also \cite{Kriele}. A further improvement was found in \cite{MI},
which contains all previous results
\item using conformal embeddings of the space-time into larger ones fulfilling the necessary assumptions \cite{Kup} and other kind of envelopments \cite{AS}
\item Theorems specially tailored for some specific situations, such as singularities in colliding plane waves \cite{T7}, see \cite{Gri} for a complete review. 
\item Theorems independent of (\ref{efe}), just assuming a metric connection and appropriate conditions such as (\ref{sec}), e.g. \cite{Low,Sir}. This is related to the study of singularity theorems in alternative theories, mainly based on rewriting the field equations in Einstein form (\ref{efe}) with an ``effective'' energy-momentum tensor, see \cite{FKLMS}.
\end{itemize}

There will be more modern singularity theorems discussed in section \ref{sec:XXI}.

Such a diversity of theorems do not have to cause too much concern to the reader. To understand their main implications and applications one can use an appropriate pattern theorem because, as argued in \cite{S1,S5}, they {\em all} share the same structural framework as condensed in the following
\begin{theorem}[Pattern Singularity Theorem]\label{th:pattern}
If a space-time of sufficient differentiability satisfies
\begin{enumerate}
\item a condition on the curvature
\item a causality condition
\item and an appropriate initial and/or boundary condition
\end{enumerate}
then it contains endless but  incomplete  causal geodesics.
\end{theorem}
The assumption  of sufficient differentiability is often ignored despite its mathematical and physical relevance. 
The singularity theorems hold if the metric $g_{\mu\nu}$ is twice differentiable with 
continuity. A breakdown of such differentiability is not a true singularity, 
specially if the first derivatives of $g_{\mu\nu}$ satisfy the Lipshitz condition. Recall that the entire metric of finite objects 
(stars or galaxies) is usually split into two different 
regions matched together at the surface of the body such that the 
$g_{\mu\nu}$ {\em cannot} be twice differentiable in GR: a jump in the matter density from the interior to the exterior is directly related to the discontinuity of the second derivatives of $g_{\mu\nu}$ via equations (\ref{efe}). As 
an example, the Oppenheimer-Snyder collapsing model (or its complementary Einstein-Straus vacuole) can not have a twice differentiable metric. Hence, this assumption is restrictive from a physical viewpoint, as actually realized, and briefly discussed, in \cite{HE}, see also \cite{Cla3,Cla4}. A list of the very many places where the differentiability assumption is used to prove the singularity theorems can be found in \cite{S1}.

\subsubsection{The condition on the curvature}\label{subsec:sec}
All singularity theorems {\em necessitate} a condition involving the Riemann tensor $R_{\alpha\beta\mu\nu}$. These are
usually presented under the names of ``energy'' and/or ``generic'' conditions but, as explained in formula (\ref{sec}), this assumption is of a {\em geometric nature}. Mostly, the used conditions are (\ref{sec}) and (\ref{gen}) or appropriate averaged versions thereof. The generic condition (\ref{gen}) is rarely violated, only in very special or symmetric situations fails. 
An interesting example is provided by the Einstein static universe: it contains trapped sets (compact slices, and points with reconverging light cones), satisfies (\ref{sec}) and is globally hyperbolic, yet it is geodesically complete. 

Condition (\ref{sec}) can be re-expressed using (\ref{efe}), but only in GR. However, (\ref{sec})  depends then on the sign of $\Lambda$ and today we know that $\Lambda> 0$, just the ÃwrongÃ
sign for the theorems. Alternative or more general theories of gravity retain the theorems, but their formulation in terms of ``energy" conditions depends on the field equations of the particular theory. 

Even in GR and independently of $\Lambda$ the curvature assumption does not always hold \cite{HE,S1} {\em classically}, such as for scalar fields and other realistic type of matter. Actually, inflationary cosmological models violate it. Notice, however, that as discussed in the previous list of improved theorems, some of them deal successfully  with this problem. For a wider discussion on energy conditions see \cite{Vi}, in relation with the theorems \cite{S1}, and for a more recent perspective \cite{Cu}. 

The curvature condition enforces the geodesic focusing via the Raychaudhuri equation (\ref{rayeq}), and thereby is absolutely indispensable.

\subsubsection{The causality condition}\label{subsec:causalcond}
The causality condition is used in two respects:
\begin{itemize}
\item to prevent the possibility of traveling to one's own past 
\item to ensure the existence of geodesics of maximal proper time between events
\end{itemize}
The first of these is not superfluous
since G\"odel's space-time contains closed 
time-like lines, but also reconverging light cones \cite{Bor2} (and minimal surfaces \cite{Ne1,Bor2,Kri,MS}), still all its geodesics are complete. As mentioned above, this part of the assumption can be substantially relaxed: it is enough that 
a region without closed timelike curves is causally separated from the set of points in such curves, see \cite{MI}.

The second point is more important, and is related to the existence of Cauchy hypersurfaces. These do not have to exist in general spacetimes. Nevertheless, in a majority of situations there will be {\em partial} Cauchy hypersurfaces: slices adapted to give initial conditions which fully determine the space-time in their domain of dependence \cite{Ge3,P5,HE}, These domains may not be the entire space-time but they are, by themselves, {\it globally hyperbolic} spacetimes. The proofs of the theorem always use such globally hyperbolic portions due to the following fundamental property: global hyperbolicity demands that the space of all causal curves between any two events is compact \cite{Le}. This is equivalent to the property that the intersection of the past of any point with the future of any other point is compact too. And then one can prove the existence of geodesics maximizing proper time between any two points of the globally hyperbolic portions of the space-time. The interesting fact here is that these maximal geodesics cannot have focal/conjugate points (and their pencils cannot have caustics), \cite{P5,HE,BE,O}. And this is the essential property used in the theorems, as explained in what follows.

\subsubsection{The boundary/initial condition}\label{subsubsec:boundcond}
Recapitulating, from the curvature condition one has
focusing of all causal geodesics, ergo the existence of 
{\em caustics and focal points} due to the focusing effect implied by the Raychaudhuri equation. 
Simultaneously, the causality condition warrants the existence of geodesics of maximal proper time, hence 
necessarily {\em without focal points}, joining pairs of events of the space-time.

A contradiction seems to twinkle --- if causal geodesics are complete. Despite appearances, however, there is no such contradiction yet. The missing clue is that we have not enforced a finite upper bound for the proper time of 
selected families of causal  geodesics. To get a contradiction with geodesic completeness one needs to add the initial/boundary condition (positive expansion of slices, closed trapped surfaces, etc.). This is why it took some time to understand how to  
combine the focusing effect with the existence of 
geodesics maximizing the arc-length to get {\em geodesic 
incompleteness}. 
The imaginative notion of closed trapped surface introduced in \cite{P} ---later generalized to compact slices and points with reconverging light cones--- eventually closed the gap.

Geodesic incompleteness cannot be proven without the initial/boundary condition. There exist globally hyperbolic spacetimes, satisfying the {\it strict} inequality (\ref{sec}) for all causal vectors, with everywhere expanding Cauchy hypersurfaces, but free of singularities \cite{S,M}. In these examples all geodesics are complete. They can avoid the conclusions of the singularity theorems because the initial/boundary condition simply fails \cite{CFS}, and therefore there is no finite bound for the maximal proper times. The focussing effect takes place fully, and at the same time there are maximal timelike geodesics, without focal points, between any two causally related events. For further details see \cite{S1,SRay}. 

Accordingly, the initial/boundary condition is absolutely essential in the theorems.

Whether or not the initial or boundary condition is satisfied by actual physical systems is debatable. 
We will probably never know if the {\em entire} Universe ---not only the observable one--- is 
spatially finite, or strictly expanding now. Thus, the formation of closed trapped surfaces in the collapse of physical systems given some realistic initial conditions has become the main area of research to elucidate this question, see end of subsection \ref{subsec:trapped2}. One can also follow a similar program, towards the past, in the case of the expansing Universe. Of course, the formation of closed trapped surfaces may depend critically on the given initial conditions, so that this is an area of active research and great relevance in numerical and mathematical relativity now, see section \ref{sec:numerics}.

\subsubsection{Ideas behind the proofs}
The geometrical basis for the proofs have been implicitly discussed in the previous paragraphs.
From a physical point of view, the proofs of the main theorems \ref{th:P}, \ref{th:HP} and their improvements can be understood, schematically, as explicated next. The boundary/initial condition
provides us with a set {\it bound to become a trapped set} (remember that a trapped set has a future (or past) with compact boundary). One of two things will happen: either the set actually becomes a trapped set, or not. In the last possibility the failure to become a trapped set implies a failure of completeness, whence
the singularity. In the former case, one may consider the whole future of the {\em compact} boundary of the future of the initial set. This is actually a future Cauchy development, which may describe the entire future or not, in the second case one knows that the boundary of this domain is {\em non-compact}. The only way that this can happen is by having endless geodesics which never reach this future boundary. Therefore, there are some geodesics that either reach a singularity or go out to infinity. 

In the last possibility the geodesic reaching infinity will have a particle horizon (boundary of its entire past) 
which is compact or reaches a singularity. The reasoning proceeds then as before
but to the past.  If again we end up  with a geodesic that comes from infinity, its combination with the first one produces the
possibility of travelling indefinitely from the past to the future, remaining
within a finite spatial region and avoiding the focusing effect. But this is
not possible if (\ref{sec}) or its averaged versions hold.

One can extract the following conclusion: under appropriate curvature conditions, 
GR favours incomplete geodesics versus trapped sets. Sometimes, this is a kind of a mystery,
for the incomplete geodesics arise even in completely empty spacetimes. Notice that
the intuitive idea that singularities must have something to do with the
existence of concentrated matter is then lost. A particular illustrative case of this situation is given by colliding wave spacetimes \cite{Gri,T7}, which only contain pure gravitational waves without matter, including realistic cases
without plane symmetry and finite wave profiles \cite{Y2}. 
One may try to argue, perhaps, that there
is a lot of gravitational energy localized in extremely small regions in this case.

\subsubsection{The conclusion of the theorems}\label{subsubsec:conclusion}
 The weakest point of the singularity theorems is their conclusions. In most 
cases this is very mild, as it can be a mere localized singularity. The theorems do not say anything, in general, about the situation, strength, extension and 
character of the singularity. One cannot even know whether it is in the 
future or past. All one knows is that there exists, at least, one incomplete causal geodesic.
Nevertheless, there are some results proving that the incomplete geodesics predicted by the theorems lead to singularities that, in general, if they are essential and not very specialized
\cite{Cla1,Cla2,Cla3}, will indicate a divergence of the curvature, see for instance
\cite{Cla1,Cla2,Cla3,Cla4,CS,Tho,TCE}.

\section{Observational consequences: cosmic censorship, critical phenomena, BKL conjecture, etc.}\label{sec:numerics}

\subsection{Cosmic censorship}\label{subsec:CC}

What are the observational consequences of the singularity theorems?  The answer to this question depends on another question: can we see singularities?  The Schwarzschild spacetime (\ref{kot}) with $\Lambda =0$ contains a singularity; but it is inside the black hole event horizon $r=\alpha$, and is therefore not visible to observers outside the black hole.  Even for observers inside the black hole, the singularity is always towards their future, and the observer cannot see the singularity before encountering it.  This leads to the question of whether gravitational collapse produces singularities that are like the singularity of Schwarzschild in these 
respects \cite{P6}, that is, are the singularities produced in gravitational collapse hidden inside black hole event horizons? (weak cosmic censorship) and are singularities non-timelike? (strong cosmic censorship).  The issue of cosmic censorship is sometimes viewed as one of the consistency of the theory, in other words that it would somehow be a disaster for general relativity if cosmic censorship were false.  However, this is probably not the right way to look at things for the following reason: general relativity is itself an approximation obtained by ignoring the quantum nature of the gravitational field.  At sufficiently large spacetime curvature (such as would be expected to occur near a spacetime singularity) this approximation breaks down.  Thus we will need quantum gravity to understand the true nature of singularities.  Cosmic censorship can thus be viewed as the question of whether the effects of quantum gravity are visible in the process of gravitational collapse, or more succinctly: do astrophysicists need to know quantum gravity?

Ideally one would like to formulate cosmic censorship as a precise mathematical conjecture and then find a proof or a counterexample.  However, this turns out to be quite difficult.  Recall that ultimately one is interested in the astrophysical process of gravitational collapse.  Therefore an overly naive formulation of a cosmic censorship conjecture might be vulnerable to what from the astrophysical point of view looks like an artificial counterexample.  For example one can easily produce ``naked singularities'' \cite{P4,PI} simply by removing regions from Minkowski spacetime \cite{GH} or by collapsing spherically symmetric dust.  Rather than regard cosmic censorship as refuted by these simple counterexamples, one refines the condition of the conjecture to rule out such counterexamples.  Of course, one could view this conjecture refining process as unfairly making cosmic censorship invulnerable to refutation by ``larding on conditions" (as one eminent astrophysicist put it).  Here the only (admittedly vague) response is that if faced with an ``astrophysically realistic" counterexample one should give up the fight rather than add a condition to rule it out.  For an extensive treatment (including a somewhat precise version) of the weak cosmic censorship conjecture see \cite{Wald1}.

One strategy to use with a conjecture is to search for a proof and also to search for a counterexample.  If a wide class of possible counterexamples can be shown to fail, then this can be regarded as evidence for the probable truth of the conjecture.  For weak cosmic censorship, one possible counterexample would be a process that turns a black hole into a naked singularity.  Recall that the Reissner-Nordstr\"om metric with (in natural units) mass $M$ and charge $Q$ represents a black hole if $Q \le M$ and a naked singularity if $Q > M$.  Similarly, the Kerr metric with mass $M$ and angular momentum $J$ represents a black hole if $J \le {M^2}$ and a naked singularity if $J > {M^2}$.  Thus, one way to produce a naked singularity would be to ``overcharge'' or ``overspin'' a black hole.  At first it might seem easy to overcharge a black hole: just shoot enough particles of charge $q$ into the black hole until finally the black hole's charge exceeds its mass.  However, it turns out that this process doesn't work: For $q$ the same sign as $Q$, the black hole will repel the charged particle.  The particle will therefore need enough energy to overcome the repulsion, and this energy will be incorporated into the mass of the black hole.  Though the black hole charge increases in this process, so does its mass, and in such a way that the charge does not exceed the mass.  Similarly, spinning black holes repel the particles whose angular momentum would increase their spin, and this ``spin-spin repulsion'' prevents the overspinning of a black hole.  

In a similar vein, Penrose showed \cite{PI} that if a certain inequality (now called the {\em Penrose inequality}) involving the area of a marginally (outer) future-trapped surface $S$ ---the ``apparent horizon''--- and the mass of the initial hypersurface containing $S$ were violated, then the spacetime that comes from evolving the initial data would contain a naked singularity. To date, the Penrose inequality has been proven only in some special cases, while no counterexample has been found either, see subsection \ref{subsec:ineq} for further details and a longer discussion.

In a globally hyperbolic spacetime, there can be no timelike singularities.  Thus, one way of formulating strong cosmic censorship is as the statement that (under suitable conditions) spacetime must be globally hyperbolic.  An initial data set has a maximal Cauchy development, which is a globally hyperbolic spacetime; however that maximal Cauchy development may not be the whole spacetime.  Thus strong cosmic censorship is the statement that generically the maximal Cauchy development is inextendible.  Here, the word ``generically'' is used because there are known special cases where the maximal development is
extendible, so without some sort of generic condition, this version of cosmic censorship would fail.  Theorems on maximal Cauchy developments come under the heading of global results for partial differential equations.  They are generally very difficult to prove (cases with spacetime symmetries are more tractable) and must be done separately for each type of matter (vacuum, Einstein-Maxwell, etc.).  For a discussion of results of this type see \cite{Ringstrom}.

\subsection{Critical gravitational collapse}

One idea for producing naked singularities comes from considering some properties of the Schwarzschild spacetime (\ref{kot}) with $\Lambda =0$.  Let $M$ be the Schwarzschild black hole mass, so that the constant $\alpha$ in (\ref{kot}) reads $\alpha =2 GM/c^2$.  Then the spacetime curvature at radius $r$ is proportional to $\alpha/{r^3}$ while the radius of the event horizon is  $\alpha$.  It then follows that the maximum curvature that is visible to outside observers is proportional to $1/\alpha^{2}$, that is, to $1/{M^2}$.  Thus the ability to produce arbitrarily small black holes is the ability to produce arbitrarily large curvature visible to outside observers.  Therefore in the limit, a configuration of matter that collapses to form a ``zero mass black hole'' actually forms a naked singularity.  Is there such a configuration of matter?  In his mathematical studies of the collapse of a spherically symmetric self-gravitating scalar field \cite{Chrcrit} Christodoulou proved the existence of just such a configuration of matter.  But what are the properties of such a ``zero mass black hole''? and what happens to configurations of the scalar field that are close to the configuration that produces it?  These questions were addressed in \cite{Ch,Chris} and nicely answered by Choptuik \cite{Choptuik} in his numerical simulations of spherically symmetric scalar field collapse.  Choptuik considered a family of initial data specified by a parameter $p$ which when evolved results in dispersal of the scalar field for small $p$ and formation of a black hole for large $p$.  There is thus some critical value $p*$ which marks the boundary between those configurations that form black holes and those that don't.  Choptuik found that for $p$ close to, but greater than $p*$ there is a scaling relation for the mass $M$ of the black hole
\begin{equation}
M \propto {(p-p*)}^\gamma
\label{Chopmass}
\end{equation} 
Furthermore there is a universality in this relation in the sense that picking a different one parameter family of initial data leads to the same scaling relation with the same value of $\gamma$.  In addition, the critical solution (the one that results from the evolution of the $p=p*$ data) has the property of discrete self-similarity: after a certain amount of time the scalar field evolves to the same profile, but with the scale of space shrunk by a factor of $e^\Delta$.  The collapse process then continues, for an infinite number of times, on an ever smaller scale of space and time, until it finally results in the formation of a naked singularity in a finite amount of proper time.  The critical solution (and the value of the constant $\Delta$) are also universal in the sense that they do not depend on which one parameter family of initial data is chosen.  Though the critical solution contains a naked singularity, the production of the naked singularity only occurs for one value of $p$ in the family and thus would not occur for generic initial data.  Thus (as for the case of strong cosmic censorship) a viable version of weak cosmic censorship takes the form that naked singularities do not result from the evolution of {\emph {generic}} initial data.   

There is a connection between the critical solution and the mass scaling relation: The critical solution has one unstable mode, which grows exponentially as $e^{\kappa \tau}$ where $\kappa $ is a constant and $\tau$ is a logarithmic time variable.  It then follows that for $p$ near $p*$ any geometric quantity with dimension of length satisfies a scaling relation of the form of eqn. (\ref{Chopmass}) with $\gamma =1/\kappa$.  In particular, since the parameter $\alpha$ in (\ref{kot}) (the black hole {\em mass} in natural units) has the dimension of length, it satisfies eqn. (\ref{Chopmass}).  Note, that this explanation of scaling requires only that $p$ is near $p*$, not that $p>p*$.  In particular, for subcritical collapse (i.e. where $p<p*$ and therefore a black hole does not form) there is some maximum value $R_{\rm max}$ achieved by the spacetime curvature during the collapse process.  Since the maximum spacetime curvature is a geometric invariant, it follows \cite{GarfDunc} that it satisfies a scaling relation of the form 
${R_{\rm max}} \propto {{({p*}-p)}^{-2\gamma}}$ where $\gamma$ is the same constant that occurs in the black hole mass scaling relation. 

Critical collapse has been studied for many forms of matter in spherical symmetry: perfect fluids \cite{Evans}, complex scalar fields \cite{Piran}, Yang-Mills fields \cite{ChopBizon} etc.  It has also been studied in axisymmetry for vacuum \cite{AE} and with scalar field matter \cite{UBCgroup}.
There is even a preliminary result for the case with no symmetry \cite{Laguna}. In all cases there is a scaling relation for black hole mass, and there is a critical solution that is either discretely self-similar or continuously self-similar.  Here the value of $\gamma$ (and $\Delta$ in the discretely self-similar case) depends on the type of matter.  For results on critical collapse, see the review \cite{Gundlach} and references therein.

\subsection{BKL singularities}\label{subsec:BKL}

As remarked in subsection \ref{subsubsec:conclusion}, one disadvantage of the singularity theorems is that they give very little information on the nature of singularities.  One normally thinks of gravitational collapse as producing an all encompassing singularity at which spacetime curvature blows up, and at which all observers who enter the black hole end their existence.  However, as explained in section \ref{sec:impact}, the singularity theorems don't say all that, merely stating that for some unspecified reason some observer or light ray ends.  

Nonetheless, there may be a simple, general description of the properties of singularities for the following reason: As with all gravitational phenomena, singularities are described by the Einstein field equations.  If curvature blows up at spacetime singularities, so should some terms in the field equations.  If terms in the field equations blow up at different rates, then a truncated set of field equations, in which only the dominant terms are kept, should give an accurate asymptotic picture of the approach to the singularity.  The trick is to guess which terms in the field equations are dominant and then to somehow verify that that guess is correct.  Such a guess was provided by Belinskii, Khalatnikov, and Lifschitz (BKL)\cite{BKL,BKL1} and usually referred to as the BKL conjecture.  BKL write the field equations in a synchronous coordinate system where time to the singularity is the time coordinate.  They then assume that the dominant terms in the field equations are those involving time derivatives (with the terms involving spatial derivatives being less important).  They then verify that their guess is consistent with the field equations.  Note that in the BKL picture, the singularity occurs at $t=0$ in a synchronous coordinate system, and is therefore spacelike, in accordance with strong cosmic censorship.  Since spatial derivatives are neglected, this means that at each spatial point the dynamics is that of a homogeneous (but anisotropic) cosmology, and that each spatial point's dynamics decouples from the dynamics of the other points, so that the singularity dynamics is local.  Furthermore, since the most general dynamics for homogeneous, anisotropic cosmologies is the oscillatory dynamics of the Bianchi type IX (Mixmaster) spacetime \cite{RSh,Exact}, BKL singularities are oscillatory.  In addition, the terms in the field equations due to the anisotropy blow up faster than those of the stress-energy (except for scalar field matter) and so the singularity is well described by the vacuum field equations.  In summary, in the BKL picture singularities are spacelike, local, and oscillatory, and ``matter doesn't matter."  But is the BKL picture right?  Objections were raised in \cite{BT} including the fact that synchronous coordinates based on a singularity cannot be assumed to be smooth, and that consistency of an asymptotic scheme with the field equations does not preclude the possibility of different asymptotic regimes that are also consistent with the field equations; nor does it mean that a general evolution towards a singularity gets itself into that asymptotic regime. 

In order to resolve these questions, Berger and Moncrief \cite{BM1} (and later several of their collaborators \cite{BM2}) embarked on a program of numerical simulations of spacetime singularities.  The idea was to pick some (fairly general) initial data, numerically evolve it towards the singularity, and see whether the evolution accorded with the BKL conjecture.  As with many research programs in numerics, one starts with situations with symmetries and then gradually works towards the general case of no symmetry.  Berger and Moncrief started with the Gowdy spacetimes \cite{Gowdy}, which have two spacelike Killing vectors.  (A special class of Gowdy spacetimes, where the Killing vectors are hypersurface-orthogonal, had been treated mathematically by Isenberg and Moncrief \cite{IM} and shown to be well described by the BKL picture). Aside from their symmetries, one feature of the Gowdy spacetimes, that makes them less challenging to study, is that they only undergo a finite number of BKL oscillations.  The Berger-Moncrief simulations showed that the BKL picture is a good description of singularities, however they also uncovered a new complication that is now called ``spikes.''  These are spatial points where the dynamics is sufficiently different from those of neighboring points to give rise to a steep spatial feature of ever narrowing width.  These spikes are a challenge to the numerics, and also a challenge to obtaining a mathematical proof that the Gowdy spacetimes satisfy the BKL conjecture.  Nonetheless, despite these difficulties, such a proof was finally obtained by Ringstr\"om \cite{Ringstrom}.

Berger and Moncrief went on to numerically simulate $U(1)$ spacetimes \cite{U1} which have only one spatial symmetry, and obtained results in agreement with the BKL picture.  Finally, the general case with no symmetries was simulated in \cite{G1} using a somewhat different method based on the scale invariant tetrad system of \cite{Uggla}.  The results of \cite{G1} agree with the BKL picture, however with not enough spatial points to resolve the spikes.  Work to improve this simulation to resolve the spikes is in progress.    

For results on the numerical studies of singularities see the review \cite{Ber} and references therein.

\subsection{Null singularities}

Though the BKL picture is consistent with the Einstein field equations, and is a good description of at least some singularities, there remains the possibility of other pictures, also consistent with the field equations and describing another class of spacetime singularities.  Such a picture was developed by Poisson and Israel \cite{ericandwerner} based on the properties of inner horizons of black holes.  The inner horizon of a Reissner-Nordstr\"om black hole is a null hypersurface and is unstable in the sense that small perturbations blow up on the horizon.  Poisson and Israel argue that these perturbations would turn the horizon into a singularity, but that it would retain its null character.  Furthermore, since the Kerr metric also has a null inner horizon, and since the black hole uniqueness theorems tell us that a general black hole settles down to the Kerr metric, the picture of \cite{ericandwerner} should be a good general description of (at least part of) the singularity inside a black hole.  The null singularity picture is supported by general arguments \cite{OF}, numerical simulations \cite{Brady} and mathematical results \cite{Daf}. If both the BKL picture and the null singularity picture are correct, then an observer who stays with the star as it is collapsing should encounter a BKL singularity, while an observer who enters the black hole long after it forms should encounter a null singularity.  

\subsection{Black strings}

The study of string theory has led to a wide consideration of the properties of general relativity in more than 4 spacetime dimensions, and in particular to the study of higher dimensional analogs of black holes.  One of the simplest such analogs is the 5 dimensional black string, which is simply the metric product of $S^1$ and the 4 dimensional Schwarzschild spacetime.  Though the Schwarzshilid metric is stable to small perturbations, it turns out the the black string is unstable \cite{GlF}.
This gives rise to the question of what is the endstate of a perturbed black string.  Based on entropic considerations, it was argued in \cite{GlF} that the black string would pinch off to become separated 5 dimensional black holes.  However it was shown in \cite{HorMae} that any such pinch off would necessitate the formation of a naked singularity.  Do black strings violate cosmic censorship?  To resolve this question, numerical simulations of the evolution of a perturbed black string were performed \cite{BS1,BS2}. The result of the simulations is that the unstable black string undergoes a cascade of formation of ever smaller spatial structures finally resulting in a pinch off in a finite amount of time.  Thus black strings form naked singularities!  We are thus left with the odd situation that the best evidence we have at this point indicates that cosmic censorship holds in 4 spacetime dimensions but is violated in 5 spacetime dimensions.

\section{Long-term impact of the theorem}\label{sec:legacy}
Many of the developments instigated by \cite{P} and the singularity theorems have been already discussed or mentioned, such as cosmic censorship, causality theory,  global hyperbolicity and Cauchy hypersurfaces, domain of dependence, conformal infinity, or black hole uniqueness theorems. The ideas behind the (proofs of the) singularity theorems have been applied to several important results in GR, such as the positive mass theorem in its original version \cite{SY1,SY2} ---as argued in \cite{Horo}---, or more recently in \cite{ChGa} (for a restricted class of spacetimes) using the Lorentzian splitting theorems \cite{Galloway}. Similarly, in \cite{AnG} the global structure of spacetimes with positive cosmological constant under the assumption of the existence of the conformal compactification is analyzed deriving strong restrictions on the topology of space by using standard arguments taken from singularity theorems.  

In this section we would like to put the emphasis on three important lines of research that, in one way or another, have their origin in the outstanding paper \cite{P} and have become branches of GR by themselves. These are: boundaries of spacetime, trapped submanifolds, and isoperimetric inequalities and the ``Hoop conjecture".


\subsection{Space-time boundaries and conformal diagrams}\label{subsec:boundary}
Singularities clearly reach, or come from, the {\em edge} of space-time. This is 
some kind of boundary which is not part of, but is accessible from within, the space-time.
 Thus the necessity of a rigourous definition of the boundary of a space-time. This boundary may even have relevance for string theory and the famous AdS/CFT correspondence, see \cite{MR,MR1}.
 
Penrose himself started this quest with the introduction of conformal infinity and conformal compactification \cite{P0,P00,P3,Fra}. The basic idea is to implant the space-time into a larger Lorentzian manifold by a {\it conformal} embedding, which ensures that the causal properties are invariant. If the original manifold obtains a boundary in the larger one, then this is defined as the conformal boundary. This allows one to treat properties of infinity, and in particular to study the properties of radiative space-times. Moreover, the boundary, being a subset of the larger manifold, inherits causal properties itself so that it acquires attributes which allow one to place it into the future or past, and to call it spacelike, null or timelike. Using these developments Penrose was able to produce explicit
expressions for the gravitational power 
radiated ``at infinity'' by an isolated system as an integral ``there" 
\cite{P3} in a fully coordinate-independent 
geometrical manner, see for further details \cite{Fra,STEWART}.

The conformal compactification can be carried out for spacetimes
such as de Sitter or the Friedman models.  This showed that singularities such as the ``creation time" are part of the boundary, and thereby they also inherit some causal properties.  In general,
however, the full conformal compactification is a chimera. Nevertheless, Friedrich has been
able to establish a procedure such that it is possible to write down a set
of equations, called the conformal field equations, that allow one to treat the problem of the existence of solutions given data at infinity as well as other relevant issues concerning conformal infinity, see e.g.  \cite{Frie,Frie0,Frie1,Frie2,Frie3} and for a review  \cite{Fra}.

Even if the full conformal compactification is not feasible, sometimes it is possible to perform the conformal
compactification of two-dimensional subsets of the space-time providing
relevant information.  This is the case of
spherically symmetric spacetimes where one can safely ignore the angular part of the space-time and deal only with the (2-dimensional) orthogonal planes. Similarly, one can concentrate on
particularly relevant 2-dimensional surfaces (ergo conformally flat) of any given spacetime.  In these cases we can draw two dimensional
pictures, called {\em Penrose diagrams} \cite{P00,P3} or {\em conformal diagrams}, obtaining information about singularities, infinity, and the global causal structure of the space-time.  A paradigmatic
example is the diagram of Kerr's space-time  axis of
symmetry, first found by Carter \cite{Ca0}, see also \cite{HE}. These extremely useful representations of spacetimes have been used largely in GR \cite{HE,Wald,S1} and, today, they are routinely drawn in many papers and extensively utilized in live discussions among scientists working on gravitation.

Once the definition of a singularity as an incomplete curve was settled, Geroch started a new line of research concerning boundaries. He introduced the {\em geodesic boundary}, or g-boundary, building equivalence classes of endless
incomplete geodesics and a notion of proximity between them \cite{Ge1}. A good property of the g-boundary is that a metric structure ---and thus some local properties of
certain singularities--- can be defined sometimes.
Unfortunately, the g-boundary has a number of problems and does not consider
non-geodesic curves, see \cite{RSh,GS} and references therein. A more complete structure is obtained with the
{\it bundle boundary} or b-boundary due to Schmidt \cite{Sc,Sc2}, based on the completion of a Riemannian metric defined on the frame bundle of the space-time. The difficulties for applying this construction are enormous and, furthermore, the cases explicitly computed have led to highly unexpected results. Yet another program was launched in 
 \cite{SS2} with the definition of the {\em abstract boundary} or a-boundary. This is a very general framework, actually 
defined for any manifold independently of having a connection or a metric, that intends to collect
every possible boundary point arising in all the envelopments of a given manifold. In the
case of spacetimes, the boundary points can be classified by using appropriate
families of curves with definite properties (such as geodesics or others),
leading to (possibly directional) singularities, points at infinity, and some
other cases \cite{SS2}. Another definition of boundary was given in \cite{GS0} combining the idea of envelopment with the original idea of conformal embedding but using the novel concept of isocausality, leading also to a generalization of conformal diagrams. This approach is more versatile than the original conformal one, but there are some subtleties \cite{FHS0} that are yet to be understood.

Probably the most important approach was put forward in \cite{GKP} with the definition of the {\em causal boundary} or c-boundary, improving on previous similar ideas that appeared in \cite{Se1}. The idea behind the c-boundary construction is to use the future or the past of endless curves which, somehow intuitively, approach either singularities or points at infinity. The set of all (indecomposable into smaller sets of the same form) past
and future sets can be considered as a completion containing all points in $V_4$ plus the
c-boundary. Some identification procedure must then be used to remove duplication of boundary points, and this may lead to great difficulties, see also \cite{HE}. The c-boundary then entered into a history of improvements too large to be detailed here, see \cite{MR,GS}, though perhaps all the issues have been finally settled, see \cite{FHS}.

For a more complete discussion on space-time boundaries up to 2005 see the review \cite{GS}, and for the more recent  advances and the latest on the c-boundary, check \cite{Flo,FH,FHS,FHS1} and references therein.

\subsection{Trapped submanifolds}\label{subsec:trapped2}
Indubitably, the most important legacy of the 1965 singularity theorem is the fundamental notion of closed trapped surface, a certainly prolific idea with many applications. It is not only very useful in the general analysis of gravitational collapse, in the formation of black holes \cite{GC1,BOM,GC2,GC3,GC4,GC5,Dafermos}, and in cosmic censorship, numerical relativity (section \ref{sec:numerics}) and isoperimetric inequalities (subsection \ref{subsec:ineq}), it has also  become an object of interest for mathematicians ---see for instance the use of trapped surfaces to prove the decay rate of gravitational radiation flux \cite{DaRo}---and it has evolved into a richer fauna of interesting ``trapped-like" submanifolds with many geometrical and physical implications. 

As remarked several times in this paper, trapped surfaces manifest themselves as having a ``wrong" causal character of a specific vector field orthogonal to the surface: ``the gradient of $r$ is timelike", where $4\pi r^2$ is the area of the surface. This actually leads to the best definition of trapped surfaces in general. We only have to identify the proper vector field whose causal character is going to determine whether or not a given surface is trapped. It turns out that this is a well-known vector field called the {\em mean curvature vector} $H^\mu$, \cite{Kri,MS,O,S2}. One virtue of this characterization is that it can be used for imbedded submanifolds of any dimension --and not only for co-dimension two, as has been traditionally the case.

For the purposes of what follows, we are going to deal with a $n$-dimensional space-time of Lorentzian signature $(-,+,\dots,+)$. Let $\zeta$ be a connected $(n-m)$-dimensional submanifold and let us denote by $\{e^\mu_A\}$ a basis of the vectors tangent to $\zeta$ ($A,B,\dots = m+1,\dots ,n$), so that the first fundamental form of $\zeta$ reads
$$\gamma _{AB}\equiv \left.g_{\mu \nu}\right|_{\zeta} e^\mu_A e^\nu_B.$$
We assume that $\gamma _{AB}$ is positive definite wherefore $\zeta$ is spacelike. 
Any $n^{\mu}$ defined on $\zeta$ and orthogonal to the tangent vectors
$$n_{\mu}e^{\mu}_A=0$$
is called a normal vector to $\zeta$. At each point on $\zeta$ there are $m$ linearly independent normal vectors. If $m>1$ all of them can be chosen to be null. The orthogonal tangent/normal splitting of the tangent spaces to $\zeta$ leads to the standard formula \cite{Kri,O}:
$$
e^\mu_A \nabla_\mu e^\nu_B =\overline{\Gamma}^{C}_{AB}e^\nu_{C}-K^\nu_{AB}.
$$
Here $\overline{\Gamma}^{C}_{AB}=\overline{\Gamma}^{C}_{AB}$ are the symbols of the Levi-Civita 
connection $\overline\nabla$ of the first fundamental form
$\gamma_{AB}$, while $K^\nu_{AB}=K^\nu_{BA}$ is called the shape tensor (or second fundamental form vector) of $\zeta$. Observe that $K^\nu_{AB}$ is orthogonal to $\zeta$. Its component along any normal $n^\mu$
$$
K_{AB}(\vec n)\equiv n_{\mu}K^{\mu}_{AB}= -n_{\mu}e^{\nu}_A\nabla_{\nu}e^{\mu}_B = e^{\mu}_Be^{\nu}_A\nabla_{\nu}n_{\mu} 
$$
is called the second fundamental form with respect to $n^{\mu}$ of $\zeta$.
The shape tensor enters in the fundamental relation
\be
e^{\mu}_{A}e^{\nu}_{B}\nabla_{\mu}v_{\nu}|_{\zeta}=\overline\nabla_{A} 
\overline{v}_{B}+v_{\mu}|_{\zeta} K^{\mu}_{AB}
\label{nablas2}
\ee
where, for all $v_{\mu}$ we denote by $\overline{v}_{B}\equiv v_{\mu}|_{S}\, e^{\mu}_{B}$
its projection to $\zeta$.

The {\em mean curvature vector} of $\zeta$ is the trace of the shape tensor  \cite{O,Kri,MS,S2}
$$
H^\mu \equiv \gamma^{AB}K^\mu_{AB}
$$
where $\gamma^{AB}$ is such that $\gamma^{AC}\gamma_{CB}=\delta^{A}_{B}$. By definition, $H^\mu$ is orthogonal to $\zeta$. Its component along any normal $n^\mu$
\be
\theta_n \equiv n_{\mu}H^\mu =\gamma^{AB}K_{AB}(\vec n) \label{expansion}
\ee
is the trace of the corresponding second fundamental form and is called the {\em expansion} of $\zeta$ along $n^\mu$.

To connect with the standard definition of trapped surface using sign of expansions, consider the traditional case of co-dimension $m=2$. Then, $\zeta$ possesses two {\em independent} normal vector fields that can be chosen to be future-pointing and null everywhere. Calling them $k^\mu_\pm $, they obey
$$
k^+_{\mu}e^\mu_{A}=0, \, \, k^-_{\mu}e^\mu_{A}=0, \, \, k^+_{\mu}k^{+\mu}=0, \, \,
k^-_{\mu}k^{-\mu}=0. 
$$
Adding a normalization condition $k_{+\mu}k_{-}^{\mu}=-1$, there still remains the boost freedom in the orthogonal plane to $\zeta$
\be
\vec{k}^+ \longrightarrow \vec{k}'^+=\sigma^2 \vec{k}, \hspace{1cm}
\vec{k}^- \longrightarrow \vec{k}'^-=\sigma^{-2} \vec{k}^- \, .\label{norm}
\ee
The mean curvature vector of $\zeta$ can then be written in this null normal basis
$$H^\mu = -\theta^- k^\mu_+ -\theta^+\,\,  k^\mu_-$$
with $\theta^{\pm}=\theta_{ k^\pm}$, called the (future) null expansions. They correspond to those previously introduced in (\ref{trapped}). Even though $\theta^\pm$ are not invariant under the boost transformations (\ref{norm}), $H^\mu$ is invariant.

The definition of (future) trapped surfaces given in (\ref{trapped}) demand that both $\theta^\pm$ are negative. This is obviously equivalent to $H^\mu$ being timelike and future directed. Consequently, one can reformulate geometrically the notion of trapped surface with the causal character of the mean curvature vector. This is sensible, because the mean curvature vector measures the variation of the area of $\zeta$ (in general, its ``$(n-m)$-volume") along directions orthogonal to $\zeta$. To fully close the relationship with the traditional cases, one should check that $H^\mu$ is simply the ``gradient of $r$" in those cases. This can be done in several ways, for instance, by using the explicit formula for $H^\mu$ given in \cite{S2} in adapted coordinate systems. 

We arrive at the definition of trapped submanifolds, and their many avatars. A spacelike submanifold $\zeta$ of {\em any dimension} is said to be 
\begin{itemize}
\item {\em future trapped}  if $H^\mu$ is timelike and future-pointing everywhere on $\zeta$, 
\item {\em weakly future trapped} if $H^\mu$ is future-pointing causal everywhere on $\zeta$ 
\item {\em marginally future-trapped}  if $H^\mu$ is future-pointing and points consistently along one of the null normals $k^\mu_\pm$ all over $\zeta$
\item {\em minimal} in the extreme case with $H^\mu=0$ everywhere on $\zeta$
\end{itemize}
 
There are always dual definitions to the past. 
For each definition there is an equivalent characterization in terms of the null expansions, see Table \ref{table:H+theta} for the salient case of co-dimension $m=2$. Observe that the extreme case where $n=m$ ($\zeta$ is a point) can be  included  somehow in the definition of trapped submanifold if the expansion along {\em every} possible null geodesic emanating from $\zeta$ becomes negative. This captures the concept of a point with reconverting light cone as appears in Theorem \ref{th:HP}.

Once we know that the causal orientation of the mean curvature vector rules whether or not a submanifold is trapped, a symbolic notation for the causal orientation of $H^\mu$ becomes very useful. Using an arrow to denote $H^\mu$ and denoting the future as the upward direction and null vectors at 45$^o$ with respect to the vertical, the symbolic notation was introduced in \cite{S0}:
\begin{center}
\begin{tabular}{c|c}
$H^\mu$ & Causal orientation \\
\hline
$\downarrow$  & past-pointing timelike\\
$\swarrow$ or $\searrow$ & past-pointing null ($\propto \vec k^+$ or $\vec k^-$) \\
$\leftarrow$ or $\rightarrow$  & spacelike\\
$\cdot$ & vanishes \\
$\nearrow$ or $\nwarrow$ & future-pointing null ($\propto \vec k^+$ or $\vec k^-$)\\
$\uparrow$ &  future-pointing timelike\\
\end{tabular}
\end{center}
The characterization using these arrows is shown in table \ref{table:H+theta} too.

\begin{center}
\begin{table}[h]
\caption{The main cases of future-trapped surfaces, characterized by the null expansions and the causal orientation of the mean curvature vector.}
{\begin{tabular}{c|c|l}
Symbol for $H^\mu$ & Expansions & Type of surface \\
\hline
$\cdot$ & $\theta^+=\theta^-=0$ &stationary or minimal \\
\hline
$\uparrow$ & $\theta^+<0, \theta^-<0$ & future trapped\\
\hline
\begin{sideways}{$\dotsearrow$}\end{sideways} & $\theta^+=0, \theta^-\leq 0$ & marginally future trapped \\
\hline
\begin{sideways}\begin{sideways}{$\dotsearrow$}\end{sideways}\end{sideways} & $\theta^+\leq 0, \theta^-=0$ & marginally future trapped \\
\hline
\begin{sideways}\begin{sideways}$\swarrow\dotsearrow$
\end{sideways}\end{sideways}\hspace{-9.9mm} \hspace{2mm}\raisebox{2mm}{$\uparrow$} & $\theta^+\leq 0, \theta^-\leq 0$ & weakly future trapped \\
\hline
\end{tabular}\label{table:H+theta}}
\end{table}
\end{center} 

In asymptotically flat situations, for example for black hole spacetimes, only the sign of the outer expansion ---that pointing towards infinity--- is relevant \cite{HE,Wald}. These can be generalized to cases where, for instance, one of the expansions is selected or favored (say because it vanishes, or has a sign). Independently of whether or not this selected direction coincides with any particular outer or external region to the surface, it has become customary in the literature to declare it to be ``outer'', and then the nomenclature speaks about ``outer trapped". See however \cite{Hay} for a more elaborated discussion. Thus, (marginally) $+$-trapped surfaces are usually referred to as (marginally) outer trapped surfaces ((M)OTS) and similarly for the `$-$' case. The main possibilities are summarized in Table \ref{ta2}.
\begin{table}[h!]
\caption{The main cases of $+$-trapped, usually called ``outer trapped'', surfaces.}
{\begin{tabular}{c|c|l}
Symbol for $H^\mu$ & Expansion & Type of surface \\
\hline
\raisebox{-1.6mm}{$\leftarrow$}\hspace{-4mm}$\unwarrow$ & $\theta^+<0$ & half converging or outer trapped (OTS)\\
\hline
& & \\
\raisebox{-2mm}{$\dotswarrow$ \hspace{-1mm}\raisebox{5.5mm}{$\nearrow$}} & $\theta^+=0$ & null dual or marginally outer trapped  (MOTS) \\ 
& & \\
\hline
$\raisebox{3mm}{$\leftarrow$} \hspace{-4.5mm} \stackrel{\begin{sideways}\begin{sideways}$\swarrow\dotsearrow$
\end{sideways}\end{sideways}\hspace{-7mm} \raisebox{2mm}{$\uparrow$}}
{\swarrow} $ &
$\theta^+\leq 0$ & weakly outer trapped (WOTS)\\
\hline
\end{tabular} \label{ta2}}
\end{table}

MOTS have been extensively studied in recent years \cite{AMS,AMS1,AM,CM}, with relevant results for black holes (in which case they are commonly called ``apparent horizons'' \cite{HE,Wald}) and the existence of marginally (outer) trapped tubes. These are hypersurfaces foliated by M(O)TS. A key result is that the boundary of the region containing OTS in a given spacelike hypersurface turns out to be a smooth MOTS \cite{KrH,AM}. There are key differences between MTSs and mere MOTSs, as the latter need an outer notion and they are usually required to enclosed a piece of a spacelike hypersurface; for examples and an enlightening discussion see \cite{Bengtsson}. In the case that the foliating surfaces are truly marginally future trapped the mentioned tubes were introduced in \cite{Hay} and called ``future trapping horizons". They are considered as good candidates to replace the event horizon of black holes \cite{H4,HE,Wald}, which happens to be teleological: the event horizon depends on future causes and is thus too globally defined ---ergo physically of little relevance \cite{AK}. The future trapping horizons, on the other hand, are defined quasilocally, hence they are more interesting to capture the concept of a (evolving or forming)  black hole that exists {\it now}. Future trapping horizons have been much studied, leading to thermodynamical properties, such as surface gravity/temperature, area/entropy law, etcetera, generalizing that of event horizons for black holes \cite{AK,Hay,Hay2,GJ,BoFa}. For their characterization from a 3+1 perspective or the physical aspects concerning them, check \cite{Jara,Nie,Booth1}. One can place restrictions on the topology of M(O)TS, classical results in this direction were proven in \cite{H4,Ne3}, and more recent results have been found in \cite{AMS,AMS1} and, in general dimension, in \cite{GaSc,Gal2}.  MOTS can also be searched for in numerical evolutions of collapsing solutions \cite{JVG,Thor}. Indeed, the presence of OTS on spatial slices is the signal used in numerical relativity to detect black holes \cite{BSh}. One also has to take care with (M)OTS when devising initial data sets \cite{Win}. 

In \cite{Hay} two versions of future trapping horizons were introduced, outer and inner. Their distinction is related to the stability of the foliating MTS, a concept first considered in \cite{Ne3} and then put on a mathematical firm basis in \cite{AMS,AMS1}, where the stability operator for MOTS was first described. This is analogous to the standard stability operator for minimal hypersurfaces in Riemannian spaces. Recall that minimal hypersurfaces extremize the area functional, and they are locally characterized by the vanishing of the mean curvature. A fundamental question concerns then the second variation of the area functional, defining the stability of the minimal hypersurface. This permits to define a self-adjoint elliptic operator on the hypersurface whose (real) spectrum governs the mentioned stability. This operator happens to be directly related to the first variation of the (vanishing) mean curvature. Marginally outer trapped surfaces, on the other hand, have a null mean curvature vector with one of the null expansions vanishing identically. They also extremize the area functional in the direction of the mean curvature vector, but the second variation is negative so that they are unstable with respect to this functional. Nevertheless, considering that in the Riemannian minimal case the operator was also associated to first variations of the vanishing mean curvature, one can consider the perturbation of the {\em vanishing expansion}; this variation leads to another elliptic stability operator analogous to the previous one {\em but} with the important difference that it is not self-adjoint in the natural sense. Despite this fact, a principal {\em real} eigenvalue can be defined and its sign rules the stability of the MOTS \cite{AMS}. A key fact here is that the second variation in the null mean curvature direction turns out to be algebraic, so that the entire variation of the vanishing expansion is ruled by a {\em unique} differential operator. This therefore translates to co-dimension two marginally outer trapped submanifolds in higher dimensions. The parallelism between MOTS and minimal hypersurfaces in Riemannian spaces can be explored further, and one can actually solve the Plateau problem for MOTS, that is, the existence of MOTS spanning a given, prescribed, boundary \cite{eichmair}.

After pioneering work by Israel \cite{Isr}, in \cite{AMS} stable MOTS are proven to belong to dynamical horizons, which are {\em spacelike} future outer trapping horizons \cite{AK,Jara}. Actually, they belong to {\em many} such horizons, one for each chosen foliation by slices in the spacetime. One ends up with a pletora of MTT, which interweave each other in very complicated ways \cite{AG}, leading to a problem of high non-uniqueness when talking about thermodynamical and other properties of locally defined black holes \cite{AK,AG,BeS,Booth}. Nevertheless, each MTT has a {\em unique} foliation by MTSs \cite{AG} ---unless the tube is null and constitutes an isolated horizon \cite{AK}, in which case any cut of this null hypersurface is a MTS. Furthermore, stable MOTS classically perpetuate as such for some time \cite{AMMS}.

Despite their local definition and their interesting quasilocal properties, closed trapped surfaces are also drastically non-local in many ways. It is known that they cannot be seen in its entirety in the Schwarzschild solution \cite{WI}, however some of them can be actually fully seen in the Oppenheimer-Snyder collapsing model, see e.g. \cite{BJS}. It might be the case that trapped surfaces associated to {\em stable} MTT are not fully visible while those associated to non-stable ones are. More dramatically, in some examples one can prove that closed trapped surfaces extend far away from the intuitive region containing the black hole, even reaching flat portions of the spacetime whose entire past is also flat \cite{BeS0}. This has been termed as the {\em clairvoyance} property of trapped surfaces \cite{BeS}, and raises the question of what is the boundary of the region in spacetime with closed future-trapped surfaces. Such an innocent-looking and seemingly simple question has turned out to be very difficult to answer: not even in spherically symmetric cases this has been solved hitherto. The new concept of {\em core} of a black hole has thereby come up: a core of black hole is a minimal region that must be removed from the spacetime in order to get rid of all closed future-trapped surfaces \cite{BeS}. In other words, this is the indinspensable region that sustains the black hole. There are some hopes that the concept of core may lead to the characterization of a preferred future trapping horizon.

As mentioned in subsection \ref{subsubsec:boundcond} the formation of closed trapped surfaces in the evolution of realistic initial data has become an important area of research. A pioneering work in this respect is \cite{GC5}, then followed by many others such as \cite{GC1} were the stability was also analyzed. In \cite{Chris1} the existence of open sets of initial data, termed ``short pulses'', leading upon evolution to the formation of future-trapped surfaces was shown rigorously. An important improvement has been recently achieved in \cite{KlRo} assuming (more general) initial conditions given on a null hypersurface. The proof requires estimates on just the first derivative of the curvature and weaker curvature controls.

Finally, the concept of closed trapped surface has also influenced the field of ``analogue gravity'' \cite{BLV}, where quantum effects in curved space-time are modeled by means of different physical systems, specially regarding the formation of ``horizons'' in fluid models. This was first considered in \cite{Unr} and much elaborated in \cite{Vis}, see \cite{BLV} for a review.

\subsection{Isoperimetric inequalities and the ``hoop" conjecture}\label{subsec:ineq}
As commented in subsection \ref{subsec:CC}, given an asymptotically flat initial data hypersurface $\Sigma$ whose total (ADM) mass is $M$, Penrose argued \cite{PI} that if the data contains an apparent horizon $S$---this is essentially a MOTS, see above---, and if the inequality 
\be
\mbox{Area}(S) \leq 16 \pi \left(GM/c^{2}\right)^{2}= \mbox{({\footnotesize in geometrized units $G=c=1$}) } 16\pi M^2 \label{PenIneq}
\ee
(now called the Penrose inequality) were violated, then the spacetime that comes from evolving the initial data would contain a naked singularity, implying a lack of predictability unknown in classical physical theories, or referring to subsection \ref{subsec:CC}, a visibility of quantum effects in macroscopic gravitational collapse. Thus initial data violating the Penrose inequality would constitute a counterexample to weak cosmic censorship, while a proof of the Penrose inequality would constitute evidence in favor of weak cosmic censorship. One can also use similar arguments if the initial data set $\Sigma$ is not asymptotically flat but rather intersects future null infinity (asymptotically hyperbolic data) by using the non-decreasing \cite{Bondi,Sa} Bondi mass for $M$. Early proofs for this case \cite{LV} were incomplete as shown in \cite{Berg}, where some advances, still inconclusive, were presented. There are some subtleties concerning the lefthand side of (\ref{PenIneq}), as pointed out in \cite{JW} and explicitly shown in \cite{Horo,Bendov}, so that actually ``Area'' there means ``the minimal area of any surface enclosing completely $S$ within $\Sigma$''. 

The heuristic argument Penrose used in \cite{PI} can be phrased in several different manners ---see for instance \cite{Mars,Malec,JVG}--- invariably invoking the singularity theorems to derive the inevitability of singularities whenever closed trapped surfaces are formed. The inequality (\ref{PenIneq}) provides a lower bound to the mass of black hole spacetimes, and thereby is related to ---and strengthens--- the positive mass theorem \cite{SY1,SY2} of general asymptotically flat spacetimes. In this sense, the rigidity part of the positive mass theorem ---the mass vanishes if and only if the initial data evolves into flat Minkowski space-time--- has a counterpart in the black hole case: equality holds if and only if the initial data originates the Schwarzschild space-time with the corresponding mass.

Neither a proof of the Penrose inequality nor a counterexample has been found in the general case, and even in spherical symmetry only a weaker version ---using the energy rather than the mass--- has been shown to hold \cite{Hay1}. However, (\ref{PenIneq}) has been proven in the so-called Riemannian case \cite{Huisken,Bray} which, from the space-time viewpoint, describes a time-symmetric situation, using the slice $\Sigma$ of time symmetry (initial data with zero extrinsic curvature), for further details see \cite{Mars,BrCh}. In this time-symmetric situation the surfaces to be used are actually minimal surfaces and the inequality (\ref{PenIneq}) is sometimes referred to as an {\em isoperimetric inequality} \cite{Gi0,Gi1}. This is reminiscent of the classical isoperimetric problem: to determine in Euclidean space a plane figure of the largest possible area whose boundary has a specified length (the perimeter). In Euclidean space this leads to the classical isoperimetric inequality $4\pi A \le L^2$ relating the length $L$ of a closed curve and the area $A$ of the planar region that it encloses, equality holding only if the curve is a circle.

The inequality (\ref{PenIneq}) has the virtue that everything depends only on the initial data set $\Sigma$, and thus it has a neat geometrical content that can be studied independently of weak cosmic censorship or of any other physical requirement. This also follows from another independent argument presented by Penrose \cite{PI} using the mass $M$ of a thin shell collapsing into flat space-time at the speed of light. This surely produces a singularity, hence under weak cosmic censorship  a black hole must form with an event horizon dressing it. One can then argue that (\ref{PenIneq}) has to hold where now $M$ is the mass of the null shell. By using energy conservation across the shell, all the quantities involved in (\ref{PenIneq}) can be computed directly {\em in flat space-time}, thus leading to a pure geometrical inequality for surfaces embedded in Minkowski space-time \cite{Gi}: the length side of the inequality is simply the integral over $S$ of its outer null expansion $\theta^{+}$. Such constructions have been analyzed in \cite{Gi,PTW,Tod,Tod0}, proving that the inequality holds for $S$ lying on constant time hyperplanes and, using a result in \cite{Tru}, this can be extended to all mean convex bodies in Euclidean space. A claim was made \cite{Gi} that this would settle the inequality for arbitrary surfaces but this cannot be true, as clearly explained in \cite{Mars}.

There exist stronger versions of the Penrose inequality involving angular momentum, electric charge, and/or the cosmological constant $\Lambda$, see \cite{Mars,Sza,JVG} and references therein. Some of these require, in order to make sense, that particular expressions under a square root are non-negative, providing some sharper versions of the positive mass theorem. In particular, the inequalities \cite{GHHP,H4,DLT,Dain}
$$
M\geq |Q| , \hspace{1cm} M\geq \sqrt{|J|}
$$
have been put forward, where $Q$ is the total electric charge of the initial data set $\Sigma$ and $J$ its total angular momentum (in geometrized units). The first of these inequalities was proven in \cite{GHHP,CRT}, while the second only makes sense if the total angular momentum is well defined, that is, in axially symmetric situations \cite{Dain,Dain2}. In this case, it has been shown to hold for vacuum and maximal initial data sets when the MTS $S$ is connected \cite{Dain1,CLW}.
This also led to the so-called area-angular momentum inequality \cite{AnP}
$$
\mbox{Area}(S) \geq 8\pi |J|
$$
which was proven in \cite{DaRe}. It also holds for totally geodesic null hypersurfaces foliated by MOTS \cite{Mars1}.  Actually, a local version for stable sections $S$ of future trapping horizons was found in \cite{JRD}, see also \cite{Simon} for the inclusion of the cosmological constant. Area inequalities involving charge can also be derived \cite{DJR,GJR}. For further details and a lengthy discussion on these matters see \cite{Dain2}.

The very same lightlike thin shell construction in flat space-time \cite{PI} discussed above has also been used \cite{PTW,Tod} to test yet another inequality concerning black holes: {\em the hoop conjecture}. This was originally formulated by Thorne in a (deliberately) vague way as \cite{Th0,MTW}
\begin{quotation}
Black hole horizons form when, and only when, a mass $M$ gets compacted into a region whose
circumference in every direction is $C\lesssim 4\pi G M /c^2$.
\end{quotation}
This statement is, to say the least, imprecise: notice the symbol $\lesssim$. For a discussion of the many difficulties and the main problems that arise when trying to give a rigorous meaning to this conjecture, see \cite{Wald1,S4} and references therein. Yet, the hoop conjecture has somehow managed to survive and, in a sense, be successful. It was settled in spherical symmetry \cite{GC4,BMO}, and discussed in more general cases in \cite{Malec}. A related mathematical result is that of \cite{SY} where upper bounds for a ÒradiusÓ of concentrated matter and lower bounds for its mass density are linked. This was used in \cite{Clar} to find conditions for the formation of future-trapped surfaces, however, examples in \cite{GC4} demonstrate that the underlying criteria are rarely met.

The main physical idea here is that black holes are extremely localized objects, so that their energy/matter content must be severely compacted in all spatial directions. Trying to make precise this idea is difficult, though. Some specific formulations were given in \cite{Fla}, and a more recent precise one in \cite{GiHoop,CGP}. A mathematical viable reformulation of the conjecture have been presented in \cite{S4}, where a long list of references can be consulted.

\section{XXI century singularity theorems}\label{sec:XXI}
Are singularity theorems something of the past? The answer to this question is a categorical `no'. The purpose of this last section is to give an idea of what is going on, of the new directions that are being explored, and of what is yet to be confirmed or improved. The whole subject has evolved and new versions, or new types, of singularity theorems are being built to take into account recent physical features such as higher dimensional theories, a (positive) cosmological constant, quantum effects, inflation, averaging, etc. 

\subsection{Mathematical advances}
To start with, and as remarked in the discussion immediately after ``Theorem" \ref{th:pattern}, it would be convenient to prove the theorems under milder differentiability assumptions, for instance if the first derivatives of the metric are Lipschitz functions, see the discussion in \cite{S1}. An important pre-requisite in that direction has been recently achieved in \cite{Ming,KSSV}, raising some optimism in this line of research. 

Recently two theorems similar to Penrose's one, but applying to infinite-space (open) cosmological spacetimes in which localized black holes may have formed were proven in \cite{VW}. The idea is to assume that a closed trapped surface lies partly outside a black-hole horizon ---something possible in a cosmological, non-asymptotically flat, context. Unlike the original theorem, the new theorems do provide some information on the location of the singularities. One conclusion is that the Universe should contain causally disconnected regions.

Comparison results concerning the area and volume of sets can be applied to re-derive singularity theorems \cite{TG}. Singularity theorems requiring the existence of an OTS, or a MOTS, rather than true (marginally) trapped surfaces have also been recently found \cite{AMMS} even considering some generalizations of the concept of MOTS \cite{EGP,CeS1}. One should also consider what happens when one or more of the hypotheses in the singularity theorems are relaxed, or suppressed altogether. This has recently been addressed in \cite{CF}, trying to find theorems with milder conclusions, and considering the ``rigidity'' part of the singularity theorems, see also \cite{GaVe}. This may open new lines worth to be explored. 

\subsection{Quantum effects}

A very important line of research arises from the tension between the singularity theorems and the (yet unfound) theory of quantum gravity. It is widely accepted that the existence of classical singularities signal a breakdown of the classical theory at extreme conditions, which is precisely when gravitational quantum effects will become relevant. Thus, there is a need to clarify if the singularity theorems, or part of them, can survive when entering into a quantum regime, or if they then simply vanish altogether. For a general discussion, see \cite{Boj}. A first step towards the analysis of singularity theorems in this respect is the weakening of the ``energy conditions'' ---also relevant in the classical regime---, that is to say, finding an appropriate version of the curvature condition in the theorems. Early results in this direction include the theorems based on averaged energy conditions \cite{T6} as discussed in subsection \ref{subsec:thms}, which were used to deal with the quantum violations of the energy conditions in \cite{Rom0}, later improved in \cite{Rom} . A larger discussion can be found in \cite{FeRo} (and references therein) and has been recently newly considered in \cite{FG}, where an analysis of Raychadhuri-like equations is performed proving that it is viable to have energy-momentum tensors which fail to satisfy even averaged energy conditions as long as an appropriate version of them ---with an exponential damping factor--- are in place. This leads to a proof of a version of the Penrose singularity theorem allowing for {\em global} violations of the energy conditions. In \cite{Ford} it was argued that one may also need to go beyond semiclassical theories and take into account the quantum fluctuations of the space-time itself, adding extra difficulties to possible quantum  singularity theorems. One problem here is that, classically, one relies on pointwise  focusing of geodesics which cannot be exactly true (despite the smallness of the fluctuations) in a quantum regime. The notion of closed trapped surface can also be generalized and adapted to quantum situations \cite{Wall}. The Penrose singularity theorem can also be proven under these weaker circumstances.

\subsection{Inflation and $\Lambda$}
There is also the question of inflation in the past of the Universe. Effectively this implies a violation of the curvature condition in the theorems, and thus one can consider the possibility that actually the Universe is past geodesically complete. This is not the case if the weak energy condition (positivity of energy density) holds \cite{Bor2,BV,BV2}, as already mentioned in subsection \ref{subsec:thms}, but as discussed in the previous paragraph the weak energy condition can be violated in inflationary models due to quantum fluctuations \cite{BV3}. This was addressed in \cite{BGV} with the result that, as long as an appropriate {\em averaged} Hubble parameter is positive, violations of the weak energy condition are not enough to avoid past incompleteness of causal geodesics.

Inflation as well as the acceleration of the expansion of the Universe are closely related to the existence of a positive cosmological constant $\Lambda >0$, which is just the wrong sign for the curvature condition (\ref{sec}) used in the focusing effect and, ultimately, in most singularity theorems. Thus the need to direct some efforts to incorporate an explicit $\Lambda >0$ in the singularity theorems. For the case of compact slices (closed case) this was studied in  \cite{Gal}, see also \cite{AnG}, proving past geodesic incompleteness under some restrictions. The classical results in \cite{Gan,Gan2,Lee} have been reconsidered in higher dimensions where the number of topological possibilities increases drastically \cite{CeS}. This is related to the topology of space \cite{CG} and the topology of M(O)TS \cite{H4,GaSc,Gal2}, and another theorem of this kind but using closed trapped circles was presented in \cite{GaS}. 

\subsection{Averages}
Both a $\Lambda\geq 0$ and {\em averaging} can be combined to obtain another form of singularity theorems for open cosmological models. As mentioned in subsection \ref{subsubsec:boundcond} there are physically acceptable globally hyperbolic {\em geodesically complete} solutions of the field equations (\ref{efe}) \cite{S,CFS,RS,S1}. Some of these solutions cannot describe the interior of finite stars: for perfect fluids this would require a timelike hypersurface with vanishing pressure, and this just does not happen in the solution \cite{S}, which also has everywhere-expanding Cauchy hypersurfaces \cite{CFS,SRay}. Thus one wonders which conditions must a space-time satisfy to be geodesically complete. For {\em stationary} globally hyperbolic spacetimes this was answered in \cite{GaH}, and the timelike component of the Ricci tensor (for the preferred Killing static observer) behaves as $\sim 1/\rho^2$ where $\rho$ is a spatial distance between any two events. This implies a fall off of the curvature for widely separated events. But, what about the dynamical, non-stationary, case? An important input into this problem was provided by Raychaudhuri himself \cite{Ray3}, who considered space-time averages of the physical quantities. He showed that non-rotating, singularity-free, open cosmological models, such as that in \cite{S}, must have vanishing space-time averages of the energy density and other relevant physical quantities. Unfortunately, this is also the case for singular spacetimes such as the standard Friedman models as was immediately noticed \cite{Saa,S3}. Nevertheless, Raychaudhuri was pointing into a very interesting direction: averaging of physical quantities. The key point is to consider {\em spatial}, rather than space-time, averages as conjectured in \cite{S3}. 

One can thus prove \cite{SRay} that if there is a non-compact Cauchy hypersurface $\Sigma$ whose expansion is positive everywhere (with asymptotic non-oscillatory behavior, a technical condition \cite{J}), the energy density and the scalar curvature on $\Sigma$ are non-negative {\em on average}, $\Lambda\geq 0$ and (\ref{sec}) holds along the geodesic vector field orthogonal to $\Sigma$, then the non-vanishing of any of the following scalars
\begin{itemize}
\item $\Lambda$
\item the averaged energy density on $\Sigma$
\item the averaged scalar curvature of $\Sigma$ 
\end{itemize}
implies that all timelike geodesics are past incomplete \cite{SRay,J}. Hence, a clear, decisive, difference between singular and
geodesically complete globally hyperbolic expanding
open cosmological models is that the latter must have a
vanishing spatial average of the matter variables.
One could thus say that any geodesically complete model is not
``cosmological'' ---if we believe that the Universe is described by
a not too inhomogenous distribution of matter.
This is, on the whole, a very satisfactory result.

\subsection{Trapped submanifolds of arbitrary dimension: extra space dimensions}
From the discussion in subsection \ref{subsec:trapped2} we know that the concept of being trapped can be associated to submanifolds of any dimension in space-time, and not only to surfaces in 4 dimensions (or co-dimension 2 submanifolds in arbitrary dimension $n$). Thus, a natural question is: why in Theorem \ref{th:P} one needs to assume a closed trapped {\em surface}? A partial answer is given by the Hawking-Penrose Theorem \ref{th:HP}, as one can assume either of
\begin{itemize}
\item a compact slice (co-dimension 1)
\item a closed trapped surface (co-dimension 2)
\item a point with reconverting light cones (co-dimension 4). 
\end{itemize}
But, what about {\em co-dimension 3}? Why closed trapped circles do not show up in this theorem? Actually, this question is even more relevant in arbitrary dimension $n$, as the number of possibilities increase. As remarked after Theorem \ref{th:HP}, the result holds in arbitrary dimension $n$ but then, why trapped submanifolds of co-dimensions $3,4,\dots ,n-1$ are missing?

This was discussed and answered in  \cite{GaS}: the outcome is that one can certainly used trapped submanifolds of {\em arbitrary} co-dimension in Theorems \ref{th:P} and \ref{th:HP} as long as the appropriate curvature condition is assumed. The key result is provided by the conditions to ensure the existence of focal points to the submanifold. Using the notation in subsection \ref{subsec:trapped2}, let $\zeta$ be the spacelike submanifold of any co-dimension $m$ and $n_{\mu}$ a future-pointing normal to $\zeta$. Let $\gamma$ represent a geodesic curve tangent to $n^\mu$ at $\zeta$ with affine parameter $u$ ($u=0$ at $\zeta$), and denote by $N^{\mu}$ the geodesic vector field tangent to $\gamma$ ($N_{\mu}|_{u=0}=n_{\mu}$).
By parallel propagating along $\gamma$ the tangent vectors $\{\vec{e}_{A}\}$ we construct a set
$\{\vec E_{A}\}$ of vector fields along $\gamma$ ($\vec{E}_{A}|_{u=0}=\vec{e}_{A}$). By construction $g_{\mu\nu}E^\mu_{A}E^\nu_{B}$ is independent of $u$, so that $g_{\mu\nu}E^\mu_{A}E^\nu_{B}=g_{\mu\nu}e^\mu_{A}e^\nu_{B}=\gamma_{AB}$. Define then $P^{\nu\sigma}\equiv \gamma^{AB}E^\nu_{A}E^\sigma_{B}$ (at $u=0$ this is the projector to $\zeta$). If the expansion (\ref{expansion}) is initally negative  $\theta_n <0$ and the curvature tensor satisfies the inequality
\be
R_{\mu\nu\rho\sigma}N^\mu N^\rho P^{\nu\sigma}\geq 0 \label{cond}
\ee
along $\gamma$, then there is a point focal to $\zeta$ along $\gamma$ at or before the affine parameter reaches the value $u=(m-n)/\theta_n$, provided $\gamma$ is defined up to that point.

It is easily checked that condition (\ref{cond}) reduces simply to (\ref{sec}) in the cases of co-dimension 1 or 2 \cite{GaS}. For co-dimension $m>2$, the interpretation of (\ref{cond}) can be given physically in terms of {\em tidal forces}, or geometrically in terms of sectional curvatures.
In physical terms, it is a statement about the attractiveness of the gravitational field on average: the tidal force in directions initially tangent to $\zeta$ is attractive on average. The Penrose singularity theorem \ref{th:P} survives as it is simply replacing the closed trapped surface for a closed trapped submanifold of arbitrary co-dimension if one uses (\ref{cond}) instead of (\ref{sec}). Actually, just a milder averaged version of (\ref{cond}) is enough \cite{GaS}. Similarly, the Hawking-Penrose theorem \ref{th:HP} holds by assuming a closed trapped submanifold of {\em any} co-dimension and (\ref{cond}).

Several applications of these generalized theorems are discussed in \cite{GaS}. Here we would just like to mention a particular one concerning the possible {\em classical} instability of spatial compactified extra dimensions. This  instability was suggested by Penrose himself in \cite{Pextra}. He argued that, due to the singularity theorems, singularities may develop within a tiny fraction of a second. His argument, though, needs some ad-hoc splittings, and some restrictions on the Ricci tensor, because theorems \ref{th:P} and \ref{th:HP} were valid only for very few co-dimensions. Those problems can be avoided by using the generalized theorems in \cite{GaS} as it is enough that the compact extra-dimensional space, or {\em any} of its compact less-dimensional subsets, satisfy the trapping condition while the restriction on Ricci curvatures can be replaced by the appropriate averaged condition on tidal forces. 
Hence, the basic argument of Penrose acquires a wider applicability and requires less restrictions.

\section{Concluding remark}
As exemplified in the previous sections, there are many exciting and interesting ideas being developed in several physical and mathematical areas which belong to the novel realm created 50 years ago by the Penrose singularity theorem. 

In conclusion, the fertile line of research engendered in \cite{P} is, today, very much alive and vibrant.

\section*{Acknowledgements}
 JMMS is supported by grants
FIS2010-15492 (MICINN), GIU12/15 (Gobierno Vasco), P09-FQM-4496 (J. Andaluc\'{\i}a---FEDER) and UFI 11/55 (UPV/EHU).
DG is supported by NSF grant PHY-1205202 to Oakland University.

\end{document}